\documentclass[aps,prd,notitlepage,nofootinbib,tightenlines]{revtex4-1}
\usepackage{amsmath}
\usepackage{bm}
\usepackage{times}
\usepackage{braket}
\usepackage{color}
\usepackage{epsfig}
\usepackage{slashed}
\usepackage{hyperref}
\newcommand{\beq}{\begin{eqnarray}}
\newcommand{\eeq}{\end{eqnarray}}
\newcommand{\ben}{\begin{enumerate}}
\newcommand{\een}{  \end{enumerate}}
\newcommand{\non}{\nonumber\\ }

\newcommand{\etar}{\eta^\prime }
\newcommand{\etap}{\eta^{(\prime)} }

\newcommand{\mbs}{m_{B_s} }

\newcommand{\psl}{ P \hspace{-2.4truemm}/ }
\newcommand{\nsl}{ n \hspace{-2.2truemm}/ }
\newcommand{\vsl}{ v \hspace{-2.2truemm}/ }

\newcommand{\cala}{ {\cal A} }
\newcommand{\calb}{ {\cal B} }

\newcommand{\calh}{ {\cal H} }
\newcommand{\ov}{ \overline  }

\def \cpc{ Chin. Phys. C  }

\def \csb{ Chin. Sci. Bull. }

\def \epjc{ Eur. Phys. J. C }

\def \npb{  Nucl. Phys. B }
\def \plb{  Phys. Lett. B }
\def \ppnp{ Prog.Part. $\&$ Nucl. Phys. }
\def \prd{  Phys. Rev. D }
\def \prl{  Phys. Rev. Lett.  }
\def \ptep{  Prog. Theor. Exp. Phys.  }

\def \zpc{  Z. Phys. C }
\def \jhep{ JHEP }

\definecolor{Red}{rgb}{1.,0.,0.}

\definecolor{Blue}{rgb}{0.,0.,1.}

\definecolor{nicered}{rgb}{0.7,0.1,0.1}
\definecolor{nicegreen}{rgb}{0.1,0.5,0.1}

\hypersetup{colorlinks,citecolor=nicegreen,linkcolor=nicered}
\begin{document}
\title{Anatomy of $B_s \to PV $ decays and effects of next-to-leading order contributions
in the perturbative QCD factorization approach}
\author{Da-Cheng Yan$^{1}$}  
\author{Ping Yang$^{1}$}   
\author{Xin Liu$^{2}$}  \email{liuxin@jsnu.edu.cn}
\author{Zhen-Jun Xiao$^{1,3}$} \email{xiaozhenjun@njnu.edu.cn}
\affiliation{$^1$ Department of Physics and Institute of Theoretical Physics,
Nanjing Normal University, Nanjing, Jiangsu 210023, P.R. China}
\affiliation{$^2$ School of Physics and Electronic Engineering, Jiangsu Normal University, Xuzhou 221116, P.R. China}
\affiliation{$^3$ Jiangsu Key Laboratory for Numerical Simulation of Large Scale Complex
Systems, Nanjing Normal University, Nanjing, Jiangsu 210023, P.R. China}
\date{\today}
\begin{abstract}
In this paper, we will make systematic calculations for the branching ratios and the CP-violating asymmetries  of the twenty one $\bar{B}^0_s \to PV $ decays by employing the perturbative QCD (PQCD) factorization approach. Besides the full leading-order (LO) contributions, all currently known next-to-leading order (NLO) contributions
are taken into account. We found numerically that:
(a) the NLO contributions can provide $\sim 40\%$ enhancement
to the LO PQCD predictions for ${\cal B}(\bar{B}_s^0  \to K^0 \bar{K}^{*0})$ and
$ {\cal B}( \bar{B}_s^0  \to K^{\pm}K^{*\mp})$, or  a $\sim 37\% $
reduction to  $ \calb(\bar{B}_s^0  \to \pi^{-} K^{*+})$;
and we confirmed that the inclusion of the known NLO contributions can improve significantly
the agreement between the theory and those currently available experimental measurements;
(b) the total effects on the PQCD predictions for the relevant $B\to P$ transition form factors
after the inclusion of the NLO twist-2 and twist-3 contributions is generally small in magnitude:
less than $ 10\%$ enhancement respect to the leading order result;
(c) for the ``tree" dominated decay $\bar B_s^0\to K^+ \rho^- $ and the ``color-suppressed-tree" decay
$\bar B_s^0\to \pi^0 K^{*0}$, the big difference between the PQCD predictions for
their branching ratios  are induced by different topological structure and by interference effects
among the decay amplitude ${\cal A}_{T,C}$ and ${\cal A}_P$: constructive for the first decay but
destructive for the second one;
and (d) for $\bar{B}_s^0 \to V(\eta, \etar)$ decays, the complex pattern of the PQCD predictions for their branching ratios can be understood by rather different topological structures and the
interference effects between the decay amplitude $\cala(V\eta_q)$ and $\cala(V\eta_s)$
due to the $\eta-\etar$ mixing.
\end{abstract}


\vspace{1cm}

\maketitle

\section{Introduction}

During the past two decades, the theoretical studies and experimental measurements
for the two-body charmless hadronic decays of $B$ and $B_s$ mesons have played a very important role in testing
the Standard Model (SM) and in  searching for the possible signals of new physics (NP) beyond SM
\cite{bfbook,lhcb0,lhcb1,lhcb2,pdg2016,hfag2016,csbbs1}.
On the theory side, such decays have been studied systematically by employing rather different
factorization approaches at the leading order (LO) or next-to-leading order (NLO), such as
the generalized factorization approach \cite{aag1,chenbs99,xiaobs01},
the QCD factorization (QCDF) approach \cite{prl99,npb675,sun2003,chengbs09} and
the perturbative QCD (PQCD) factorization approach \cite{li2003,nlo05,pqcd1,pqcd2,joint}.
The resultant theoretical predictions from different approaches
are generally consistent with each other within the errors.

On the experimental side, the early measurements for $B \to M_2 M_3$ decay modes
( here $M_i$ stands for the light pseudo-scalar or vector mesons ) mainly
come from the BaBar and Belle collaboration in B factory experiments \cite{bfbook,hfag2016}.
For  $B_s \to M_2 M_3$ decays, however, LHCb Collaboration provide the dominant
contribution \cite{lhcb0,lhcb1,lhcb2,pdg2016,hfag2016}. Although some deviations or
puzzles, such as the so-called $(R(D),R(D^*))$ and $(R_K,R_{K*})$ anomalies,
are observed so far, but there is no any solid flavor-related evidence for the existence of the new physics
beyond the SM.

In the framework of the PQCD factorization approach, the charmless two-body hadronic decays
$B_s^0 \to M_2 M_3$ have been studied by some authors in recent years:
\begin{enumerate}
\item[(1)]
In 2004, Li et al. studied the pure annihilation $B_s \to \pi^+\pi^-$
decay \cite{bspipi} and gave a leading order PQCD prediction for a large
branching ratio ${\cal B}(B_s^0 \to \pi^+\pi^-) \sim 5\times 10^{-7}$, which has been
confirmed by recent CDF and LHCb measurements \cite{cdf1,lhcb1a,lhcb1b,xiao2012}.

\item[(2)]
In 2007, Ali et al. completed the systematic study for the forty-nine
$B_s^0 \to PP, PV, VV$ decays at the LO level, presented their
PQCD predictions for the CP-averaged branching ratios, the CP-violating asymmetries and some other
physical observables~\cite{ali07} .
For $B_s\to \pi^+\pi^-$, for example, they also found a large theoretical prediction for its decay rate.

\item[(3)]
In 2014, Qin et al. studied the twenty $B_s \to P T$ decays ( here $P$ and $T$ denote the light pseudo-scalar
and tensor mesons ) in the PQCD factorization approach at the LO level,
and provided their predictions for the decay rates and CP-violating asymmetries of those considered decay modes
\cite{qin2014}.

\item[(4)]
Very recently, we studied $ B^0_s \to (K\pi,KK) $ decays \cite{xiao14a} and $B_s^0\to (\pi \etap, \etap\etap)$ decays
\cite{xiao14b} at the partial NLO level. We found that the currently known NLO contributions
from different sources can interfere with the LO part constructively or destructively for different
decay modes, while the agreement between the central values of the PQCD predictions for the decay rates
and CP violating asymmetries and those currently available experimental measurements
are  indeed improved effectively after the inclusion of those NLO contributions \cite{xiao14a,xiao14b}.

\end{enumerate}

In this paper, by employing the PQCD factorization approach, we will make a systematic study for
all two-body charmless hadronic decays $B_s \to PV$  ( here $P=(\pi, K, \eta,\etar)$ and $V=(\rho,K^*, \phi, \omega)$ ),
by extending the previous LO studies to the partial NLO level: including all currently known NLO contributions.
We will focus on investigating the effects of the NLO contributions, specifically those  newly known
NLO twist-2 and twist-3 contributions to the form factors of $B \to P$ transitions \cite{prd85-074004,cheng14a}
under the approximation of $SU(3)$ flavor symmetry.

This paper is organized as follows. In Sec.~\ref{sec:lo-nlo}, we give a brief review about the PQCD factorization approach
and we calculate analytically the relevant Feynman diagrams and present the various decay amplitudes
for the considered decay modes in the LO and NLO level.
We calculate and show the PQCD predictions for the branching ratios and  CP violating asymmetries of all
twenty-one $B_s \to PV $ decays in Sec~\ref{sec:n-d}. The summary and some discussions are included in
Sec.~\ref{sec:4}.

\section{ Decay amplitudes at LO and NLO level}\label{sec:lo-nlo}

As usual, we consider the $B_s$ meson at rest and treat it as a heavy-light system.
Using the light-cone coordinates, we define the $B_s^0$ meson with momentum
$P_1$, the emitted meson $M_2$ and the recoiled meson $M_3$ with momentum $P_2$
and $P_3$ respectively. We also use $x_i$ to denote the momentum fraction
of anti-quark in each meson and set the momentum $P_i$ and $k_i$ ( the momentum carried by the
light anti-quark in $B_s$ and $M_{2,3}$ meson) in the following forms:
\beq
P_1 &=& \frac{\mbs}{\sqrt{2}} (1,1,{\bf 0}_{\rm T}), \quad
P_2 = \frac{M_{B_s}}{\sqrt{2}}(1,0,{\bf 0}_{\rm T}), \quad
P_3 = \frac{M_{B_s}}{\sqrt{2}} (0,1,{\bf 0}_{\rm T}),\non
k_1 &=& (x_1 P_1^+,0,{\bf k}_{\rm 1T}), \quad
k_2 = (x_2 P_2^+,0,{\bf k}_{\rm 2T}), \quad
k_3 = (0, x_3 P_3^-,{\bf k}_{\rm 3T}).
\eeq
The integration over $k_{1,2}^-$ and $k_3^+$  will lead conceptually to the
decay amplitude
\beq
\cala \sim \int\!\! d x_1 d x_2 d x_3 b_1 d b_1 b_2 d b_2 b_3 d b_3 \cdot
\mathrm{Tr}\left [ C(t) \Phi_{B_s}(x_1,b_1) \Phi_{M_2}(x_2,b_2) \Phi_{M_3}(x_3, b_3) H(x_i,
b_i, t) S_t(x_i)\, e^{-S(t)} \right ], \quad \label{eq:a2}
\eeq
where $b_i$ is the conjugate space coordinate of $k_{\rm iT}$,
$C(t)$ are the Wilson coefficients evaluated at the scale $t$,
and $\Phi_{B_s}$ and $\Phi_{M_i}$ are wave functions of the $B_s$ meson and
the final state mesons. The hard kernel $H(x_i,b_i,t)$ describes the four-quark
operator and the spectator quark connected by a hard gluon.
The Sudakov factor $e^{-S(t)}$ and $S_t(x_i)$ together suppress the soft
dynamics effectively \cite{li2003}.

\subsection{ Wave functions and decay amplitudes}\label{sec:wf}

For the considered $\bar{B}^0_s \to P V$ decays with a quark level transition $b \to q$ with $q=(d,s)$,
the weak effective Hamiltonian $H_{eff}$ can be written as\cite{buras96}
\beq
\label{eq:heff}
{\cal H}_{eff} &=& \frac{G_{F}}{\sqrt{2}}     \Bigg\{ V_{ub} V_{uq}^{\ast} \Big[
 C_{1}({\mu}) O^{u}_{1}({\mu})  +  C_{2}({\mu}) O^{u}_{2}({\mu})\Big]
  -V_{tb} V_{tq}^{\ast} \Big[{\sum\limits_{i=3}^{10}} C_{i}({\mu}) O_{i}({\mu})
  \Big ] \Bigg\} + \mbox{h.c.}
\eeq
where $G_{F}=1.166 39\times 10^{-5}$ GeV$^{-2}$ is the Fermi constant, and
$V_{ij}$ is the Cabbibo-Kobayashi-Maskawa (CKM) matrix element, $C_i(\mu)$ are the Wilson coefficients and $O_i(\mu)$
are the four-fermion operators.

For $B_s^0$ meson, we consider only the contribution of Lorentz structure
\beq
\Phi_{B_s}= \frac{1}{\sqrt{6}} (\psl_{B_s} +m_{B_s}) \gamma_5 \phi_{B_s} ({\bf k_1}),
\label{eq:bsmeson}
\eeq
and adopt the distribution amplitude $\phi_{B_s}$ as in Refs.~\cite{bspipi,ali07,xiao14a}.
\beq
\phi_{B_s}(x,b)&=& N_{B_s} x^2(1-x)^2 \exp \left  [ -\frac{M_{B_s}^2\ x^2}{2 \omega_{B_s}^2}
-\frac{1}{2} (\omega_{B_s} b)^2\right].
\label{phib}
\eeq
We also take $\omega_{B_s} =0.50 \pm 0.05$ GeV in numerical calculations.
The normalization factor $N_{B_s}$ will be determined  through the
normalization condition: $\int_0^1 dx \; \phi_{B_s}(x,b=0)=f_{B_s}/(2\sqrt{2N_c})$.

For $\eta$-$\etar$ mixing, we also use the quark-flavor basis:
$\eta_q= (u\bar u +d\bar d)/\sqrt{2}$ and $\eta_s=s\bar{s}$ \cite{fks98,xiao08b}.
The physical $\eta$ and $\etar$ can then be written in the form of
\beq
\left(\begin{array}{c} \eta \\ \eta^{\prime}\end{array} \right)= \left ( \begin{array}{cc}
\cos\phi & -\sin\phi\\ \sin\phi & \cos\phi\\ \end{array} \right)
\left(\begin{array}{c} \eta_q \\ \eta_s\end{array} \right),\label{eq:e-ep2}
\eeq
where $\phi$ is the mixing angle. The relation between the decay constants
$(f_\eta^q, f_\eta^s,f_{\etar}^q,f_{\etar}^s)$ and $(f_q,f_s)$ can be found for example
in Ref.~\cite{xiao08b}. The chiral enhancements $m_0^{\eta_q}$ and $m_0^{\eta_s}$ have been defined
in Ref.~\cite{li2006} by assuming the exact isospin symmetry $m_q=m_u=m_d$.
The three input parameters $f_q, f_s,$ and $\phi$ in Eq.~(\ref{eq:e-ep2})
have been extracted from the data \cite{fks98}
\beq
f_q=(1.07\pm 0.02)f_{\pi},\quad f_s=(1.34\pm 0.06)f_{\pi},\quad \phi=39.3^\circ \pm 1.0^\circ.
\eeq
With $f_\pi=0.13$ GeV, the chiral enhancements $m_0^{\eta_q}$ and $m_0^{\eta_s}$ consequently
take the values of $m_0^{\eta_q}=1.07$ GeV and $m_0^{\eta_s}=1.92$ GeV \cite{li2006}.

For the final state pseudo-scalar mesons $M=(\pi, K,\eta_q,\eta_s)$, their wave functions are the same
ones as those in Refs.~\cite{pball98,pball06,csbwf,fan2013}:
\beq
\Phi_{M_i}(P_i,x_i)\equiv \frac{1}{\sqrt{6}}\gamma_5 \left [ \psl_i \phi^{A}_{M_i}(x_i)
+m_{0i} \phi_{M_i}^{P}(x_i)+ \zeta m_{0i} (\nsl \vsl -1)\phi_{M_i}^{T}(x_i)\right ],
\label{eq:phip}
\eeq
where $m_{0i}$ is the chiral mass of the meson $M_i$,  $P_i$ and $x_i$ are the momentum and
the fraction of the momentum of $M_i$s. The parameter $\zeta=1$ or $-1$ when the
momentum fraction of the quark (anti-quark) of the meson is set to be $x$.
The distribution amplitudes (DA's) of the meson $M$ can be found easily in Refs.~\cite{bspipi,fan2013}:
\beq
\phi_{M}^A(x) &=&  \frac{3 f_M}{\sqrt{6} } x (1-x)
    \left[1+a_1^{M}C^{3/2}_1(t)+a^{M}_2C^{3/2}_2(t)+ a_4^M C_4^{3/2}(t) \right],\label{eq:piw1}\\
\phi_M^P(x) &=&   \frac{f_M}{2\sqrt{6} }
   \left \{ 1+\left (30\eta_3-\frac{5}{2}\rho^2_{M} \right ) C^{1/2}_2(t)
   -3\left [ \eta_3 \omega_3 + \frac{9}{20}\rho_M^2\left ( 1 + 6a_2^M\right)C_4^{1/2}(t)\right]
   \right \}, \ \
\label{eq:piw2}   \\
\phi_M^T(x) &=&  \frac{f_M(1-2x)}{2\sqrt{6} }
   \left\{ 1+6\left [ 5\eta_3-\frac{1}{2}\eta_3\omega_3-\frac{7}{20}\rho^2_M
   -\frac{3}{5}\rho^2_M a_2^{M} \right ]
   \left (1-10x+10x^2\right )\right \},\quad
   \label{eq:piw3}
\eeq
where $t=2x-1$, $f_M$ and $\rho_M$ are the decay constant and  the mass ratio with the definition of
$\rho_M=(m_\pi/m_0^\pi,m_K/m_0^K$, $m_{qq}/m_0^{\eta_q},m_{ss}/m_0^{\eta_s})$. The parameter $m_{qq}$ and $m_{ss}$ have been
defined in Ref.~\cite{li2006}:
\beq
m_{qq}^2 &=& m_\eta^2 \cos^2\phi + m_{\etar}^2 \sin^2\phi
- \frac{\sqrt{2}f_s}{f_q} (m_{\etar}^2-m_\eta^2) \cos\phi \sin\phi , \non
m_{ss}^2 &=& m_\eta^2 \sin^2\phi + m_{\etar}^2 \cos^2\phi
- \frac{\sqrt{2}f_q}{f_s} (m_{\etar}^2-m_\eta^2) \cos\phi \sin\phi ,
\eeq
with the assumption of exact isospin symmetry $m_q=m_u=m_d$.
The explicit expressions of those Gegenbauer polynomials $C_1^{3/2}(t)$ and $C_{2,4}^{1/2,3/2}(t)$ can be found
for example in Eq.~(20) of Ref.~\cite{xiao08b}.
The Gegenbauer moments $a_i^M$ and other input parameters are the same as those in Ref.~\cite{pball06}
\beq
a^{\pi,\eta_q,\eta_s}_1 &=& 0,\quad a^K_1  = 0.06, \quad a^{\pi,K}_2=0.25\pm 0.15, \quad a^{\eta_q,\eta_s}_2=0.115, \non
a^{\pi, K,\eta_q,\eta_s}_4  &=& -0.015, \quad \eta_3= 0.015, \quad \omega_3=-3.0.
\label{eq:aim}
\eeq

For the $\bar{B}^0_s \to P V$ decays, only the longitudinal polarization component
of the involved vector mesons contributes to the decay amplitude.
Therefore we choose the wave functions of the vector mesons as in Ref.~\cite{ali07}:
\beq
\Phi_{V}^{||}(P,\epsilon_L)\equiv \frac{1}{\sqrt{6}}
\left [\not\! \epsilon_L M_V\phi_{V}(x)+\not\! \epsilon_L\psl\phi^{t}_{V}(x)
+M_V\phi^{s}_{V}(x)\right],
\label{eq:phiv}
\eeq
where $P$ and $M_V$ are the momentum and the mass of the light vector mesons,
and  $ \epsilon_L $ is the longitudinal polarization vector of the vector mesons.
The twist-2 distribution amplitudes $\phi_V(x)$ in Eq.~(\ref{eq:phiv})
can be written in the following form~\cite{ali07}
\begin{eqnarray}
\phi_\rho (x)&=&\frac{3f_\rho}{\sqrt{6}} x (1-x)\left[1+ a_{2\rho}^{||}C_2^{3/2} (t) \right]\;,\label{phirho}\\
\phi_\omega(x)&=&\frac{3f_\omega}{\sqrt{6}} x (1-x)\left[1+ a_{2\omega}^{||}C_2^{3/2} (t)\right]\;,\\
\phi_{K^*}(x)&=&\frac{3f_{K^*}}{\sqrt{6}} x (1-x)\left[1+a_{1K^*}^{||}C^{3/2}_1(t)+
a_{2K^*}^{||}C_2^{3/2} (t)\right]\;,\\
\phi_{\phi}(x)&=&\frac{3f_{\phi}}{\sqrt{6}} x (1-x)\left[1+ a_{2\phi}^{||}C_2^{3/2} (t)\right]\;,\label{phiphi}
\end{eqnarray}
where $t=2x-1$, $f_{V}$ is the decay constant of the vector meson
with longitudinal  polarization.
The Gegenbauer moments here are  the same as those in Ref.~\cite{ali07}:
\begin{eqnarray}
a_{1K^*}^{||}=0.03\pm0.02,\quad
a_{2\rho}^{||}=a_{2\omega}^{||}=0.15\pm0.07,\quad
a_{2K^*}^{||}=0.11\pm0.09,\quad  a_{2\phi}^{||}=0.18\pm0.08.
\end{eqnarray}
While the twist-3 distribution amplitudes $\phi^t_V(x)$ and $\phi^s_V(x)$
are defined with the asymptotic form as in Ref.~\cite{ali07}:
\begin{eqnarray}
\phi^t_V(x) = \frac{3f^T_V}{2\sqrt 6}(2x-1)^2,
  \quad \phi^s_V(x)=\frac{3f_V^T}{2\sqrt 6} (1-2x)~,
\end{eqnarray}
where $f_{V}^T$ is the decay constant of the vector meson with transverse  polarization.

\subsection{ Example of the LO decay amplitudes}\label{sec:lo-aml}

At the LO level, the twenty one $\bar{B}^0_s \to P V$ decays have been studied
previously in Ref.~\cite{ali07}, and the decay amplitudes as presented in Ref.~\cite{ali07}
are confirmed by our independent recalculation.
In this paper, we  focus on  the calculations of the effects of all currently known
NLO contributions to these  decay modes in the PQCD factorization approach.
The relevant Feynman diagrams  which may contribute
to the considered $\bar{B}_s^0$ decays at the leading order are illustrated in Fig.1.

\begin{figure}[tb]
\vspace{-5cm}
\centerline{\epsfxsize=18cm \epsffile{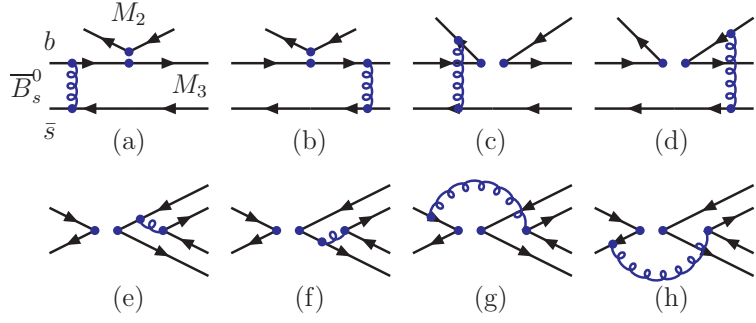}}
\vspace{-16cm}
\caption{ The typical Feynman diagrams which may contribute at leading order to $\bar{B}^0_s \to PV$ decays:
(a) and (b) are the factorizable emission diagrams; (c) and (d) the hard-spectator diagrams;
and (e)-(h) annihilation diagrams.}
\label{fig:fig1}
\end{figure}

Based on the effective Hamiltonian ${\cal H}_{eff}$, each considered decays may receive
contributions from one or more terms proportional to different Wilson coefficients $C_i(\mu)$ and/or their combinations
$a_i$ \footnote{For the sake of simplicity, one usually define the combinations of the Wilson coefficient in the
form:
$a_1=C_2+C_1/3$, $a_2=C_1+C_2/3$, $a_i=C_i+C_{i+1}/3$ and
$a_j=C_j+C_{j-1}/3$ for $i=(3,5,7,9)$ and $j=(4,6,8,10)$. For given $\mu =[2, 5]$ GeV, one found numerically
\cite{ali07,zhang09}:
$a_1\approx C_2 \approx 1.1$ are large quantity, $C_1\sim -0.2$ and $a_2=0.01-0.1$ are small ones,
the QCD-penguins $|a_{3-6}|=0.01-0.001$ are very small, and finally  the electroweak-penguins
$|a_{7-10}|=10^{-3}-10^{-4}$ are indeed tiny.  }.
According to the topological structure of the relevant Feynman diagrams for a given decay mode,
i.e.  which diagram provides the dominant contribution, one can classify the decays considered
into the following four types:
\ben
\item[(1)]
The ``color-allowed-tree" (``${\bf\rm T}$") decay: the dominant contribution comes
from the terms proportional to $a_1$ and/or $C_2$;

\item[(3)]
The ``color-suppressed tree" (``${\bf\rm C}$") decay:  the terms with
$a_2$ and/or $C_1$ provide the dominant contribution;

\item[(3)]
The ``QCD penguin" (``${\bf\rm P}$") decay and the ``Electroweak penguin" (``${\bf \rm P_{\rm EW}}$")
decays: the dominant terms are proportional to $C_{3-6}$ or $a_{3-6}$ and $C_{7-10}$ or $a_{7-10}$, respectively;

\item[(4)]
The ``annihilation" (``${\bf\rm Anni}$") decays: if only the annihilation diagrams contribute.
\een

At the leading order PQCD approach, as illustrated in Fig.~\ref{fig:fig1},
there are three types of diagrams contributing to the $\bar{B}^0_s \to PV$
decays considered in this paper: the factorizable emission diagrams ( Fig.~\ref{fig:fig1}(a) and \ref{fig:fig1}(b));
the hard-spectator diagrams (Fig.~\ref{fig:fig1}(c) and \ref{fig:fig1}(d));
and the annihilation diagrams (Fig.~\ref{fig:fig1}(e)-\ref{fig:fig1}(h)).
From the factorizable emission diagrams Fig.~\ref{fig:fig1}(a) and \ref{fig:fig1}(b),
the corresponding form factors of $B_s \to M_3$ transition
can be extracted by perturbative calculations.

For the sake of completeness and the requirement for later discussions, we show here the total LO decay amplitudes
for $\bar{B}^0_s \to\pi^{-}K^{*+}$,   $ K^+\rho^{-}$ and $ \pi^0 K^{*0}$ decays.
For other eighteen decay modes, one can found the expressions of their LO decay amplitudes easily
in Ref.~\cite{ali07}.
\beq
 A(\bar B_{s}^0\to\pi^{-}K^{*+}) &=& V_{ub}V_{ud}^{*}\Big\{ f_{\pi}
F_{eK^*}\; a_1  + M_{eK^*} \; C_{1}
\Big \}-V_{tb}V_{td}^{*} \Bigg\{ f_{\pi}F_{e K^*}\left[   a_{4} +a_{10}\right]  \nonumber  \\
&&-f_{\pi} F_{e K^*}^{P_2} \left[a_{6} +a_{8}\right]
 +  M_{e K^*} \left[C_{3}+C_{9}\right]  + f_{B_s}
 F_{a K^*}\left[a_{4}-\frac{1}{2}a_{10}\right]\nonumber  \\
&&- f_{B_s} F_{a K^*}^{P_2}\left[a_{6}- \frac{1}{2}a_{8}\right]
  + M_{a K^*}\left[C_{3}-\frac{1}{2}C_{9}\right] - M_{a K^*}^{P_1}\left[C_{5}-\frac{1}{2}C_{7}\right]
\Bigg \}, \label{eq:a001}\\
A(\bar B_{s}^0\to K^+\rho^{-})&=&V_{ub}V_{ud}^{*} \Big\{ f_{\rho}
F_{e K}\; a_1 + M_{e K} \; C_{1} \Big\} -V_{tb}V_{td}^{*} \Bigg\{ f_{\rho}F_{e K}\left[   a_{4} +a_{10}\right]  \non
&&+  M_{e K} \left[C_{3}+C_{9}\right]
+ M_{eK}^{P_1}\left[C_{5}+C_{7}\right] + f_{B_s}  F_{aK}\left[a_{4}-\frac{1}{2}a_{10}\right] \non
 &&+ f_{B_s} F_{aK}^{P_2}\left[a_{6}-\frac{1}{2}a_{8}\right]
 + M_{aK}\left[C_{3}-\frac{1}{2}C_{9}\right] + M_{aK}^{P_1}\left[C_{5}-\frac{1}{2}C_{7}\right] \Bigg \},
\label{eq:a002}\\
\sqrt{2}A(\bar B_{s}^0\to\pi^{0} K^{*0}) &=& V_{ub}V_{ud}^{*} \Big \{f_{\pi} F_{e
K^*} \; a_{2} + M_{e K^*}\; C_{2}\Big \}  -V_{tb}V_{td}^{*} \Bigg \{f_{\pi} F_{e K^*} \left[
 -a_{4}-\frac{3}{2}a_7+\frac{1}{2}a_{10}+\frac{3}{2}a_9\right] \non
&&\hspace{-1cm}  - f_{\pi} F_{eK^*}^{P_2}   \left[-a_{6}+\frac{1}{2}a_{8}\right]
 +  M_{eK^*}   \left[-C_{3}+\frac{3}{2}C_{8}+\frac{1}{2}C_{9}
  +\frac{3}{2}C_{10}\right]    -f_{B_s} F_{aK^*}^{P_2}\left[-a_{6}+\frac{1}{2}a_{8}\right]   \non
&& \hspace{-1cm}+  f_{B_s}  F_{aK^*}\left[-a_{4}+ \frac{1}{2}a_{10}\right]
  +  M_{aK^*}\left[-C_{3}+\frac{1}{2}C_{9}\right]    -  M_{aK^*}^{P_1}\left[-C_{5}+\frac{1}{2}C_{7}\right]\Bigg \},
\label{eq:a003}
\eeq
where $a_i$  are the combinations of the Wilson coefficients $C_i$ \cite{ali07}.
The individual decay amplitudes appeared in the above equations, such as $F_{eM_3}, F_{eM_3}^{P2},M_{eM3},F_{aM_3}$
and $M_{aM3}$, are obtained by evaluating the Feynman diagrams in Fig.~\ref{fig:fig1} analytically.
The term $F_{eM_3}$ and $F_{eM_3}^{P2}$, for example, comes from the factorizable emission diagrams
with $(V-A)(V-A)$ and $(S-P)(S+P)$ current, respectively.

The explicit expressions of $F_{eM_3}$ and other decay amplitudes at the leading order in PQCD approach
can be found, for example, in Ref.~\cite{ali07}.
For the sake of the conveniens of the reader, we show $F_{eP}$, $F_{eV}$ and $F_{eV}^{P_2}$ here explicitly:
\begin{eqnarray}
F_{e P} &=& 8\pi C_FM_{B_s}^4 f_V \int^1_0dx_1dx_3\int^\infty_0b_1db_1b_3db_3
\; \phi_{B_s}(x_1,b_1) \non
&& \times  \Big\{ \Big [ (1+x_3)\phi_p^A(x_3)+r_p(1-2x_3)(\phi_p^P(x_3)+\phi_p^T(x_3))  \Big ]
\cdot \alpha_s(t_a) E_e(t_a) h_e(x_1,x_3,b_1,b_3)\non
&& \ \ \ \ + 2r_p \phi_p^P(x_3)\cdot \alpha_s(t_b) E_e(t_b) h_e(x_3,x_1,b_3,b_1) \Big\},\label{ppefll}
\eeq
\beq
F_{e V} &=&8\pi   C_FM_{B_s}^4 f_P \int^1_0dx_1dx_3\int^\infty_0b_1db_1b_3db_3\; \phi_{B_s}(x_1,b_1) \non
&& \times  \Big\{ \Big[ (1+x_3)\phi_v(x_3)+r_v(1-2x_3)\left [\phi_v^s(x_3)+\phi_v^t(x_3) \right]
\Big]\cdot \alpha_s(t_a) E_e(t_a) h_e(x_1,x_3,b_1,b_3) \non
&& \ \ \ \ + 2r_v\phi_v^s(x_3)\cdot \alpha_s(t_b) E_e(t_b)h_e(x_3,x_1,b_3,b_1)  \Big\},\label{pvefll}
\eeq
\beq
F^{P_2}_{eV}&=& 16\pi r_p  C_FM_{B_s}^4 f_P \int^1_0dx_1dx_3\int^\infty_0b_1db_1b_3db_3 \; \phi_{B_s}(x_1,b_1)\non
&& \times \Big\{ \Big[\phi_v(x_3)+r_v(2+x_3)\phi_v^s(x_3)-r_vx_3\phi_v^t(x_3)\Big]\cdot \alpha_s(t_a)E_e(t_a)
  h_e(x_1,x_3,b_1,b_3)\non
&&\ \ \ \ +2   r_v\phi_v^s(x_3)\cdot \alpha_s(t_b)E_e(t_b)h_e(x_3,x_1,b_3,b_1)\Big\},
\eeq
where $C_F=4/3$ and $\alpha_s(t_i)$ is the strong coupling constant. In the above functions,
$r_v=M_v/M_{B_{s}}$ and $r_p=m_0^{P}/M_{B_{s}}$ with $m_0^P$ the chiral mass of the
pseudoscalar meson. The explicit expression of the functions $E_i(t_j)$, the
hard scales $t_i$, the hard functions $h_i(x_j,b_j)$ and more details
about the LO decay amplitudes can also be found in Ref.~\cite{ali07}.

\subsection{ NLO contributions}

\begin{figure}[tb]
\vspace{-5cm}
\centerline{\epsfxsize=18 cm \epsffile{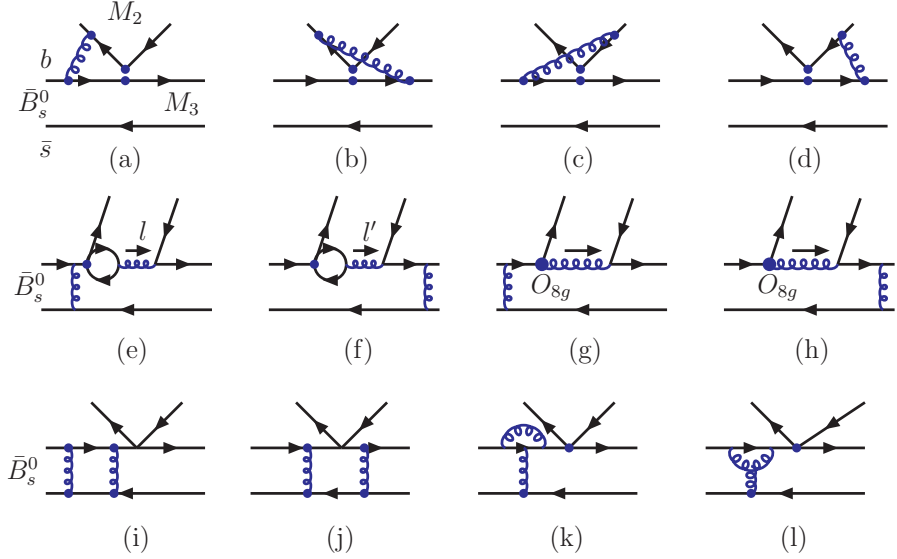}}
\vspace{-13cm}
\caption{Typical Feynman diagrams for NLO contributions:  the vertex corrections (a-d);
the quark-loops (e-f),  the chromo-magnetic penguin contributions (g-h),
and the NLO twist-2 and twist-3 contributions to $B_s \to (K, \eta_s)$ transition form factors (i-l).}
\label{fig:fig2}
\end{figure}

During the past two decades, many authors have made great efforts to calculate the NLO contributions to
the two-body charmless decays $B/B_s\to M_2 M_3$ in the framework of the PQCD factorization approach.
At present, almost all such NLO contributions become available now:
\begin{enumerate}
\item[(1)]
The NLO Wilson Coefficients (NLO-WC): which means that the NLO Wilson coefficients $C_i(m_W)$,
the renormalization group running matrix $U(m_1,m_2,\alpha)$ at NLO level ( for details see Eq.~(7.22) of
Ref.~\cite{buras96}) and the strong coupling constant $\alpha_s(\mu)$ at two-loop level will be used
in the numerical calculations \cite{buras96}, instead of the ones at the LO level.

\item[(2)]
The NLO vertex corrections (VC) as given in Refs.~\cite{npb675,nlo05}, and as illustrated
in Fig.~\ref{fig:fig2}(a)-\ref{fig:fig2}(d).

\item[(3)]
The NLO contributions from the quark-loops (QL) as described in Ref.~\cite{nlo05}, with the relevant
Feynman diagrams as shown in Fig.~\ref{fig:fig2}(e) and \ref{fig:fig2}(f).

\item[(4)]
The NLO contributions from the chromo-magnetic penguin (MP) operator $O_{8g}$ \cite{o8g2003}, as illustrated
in Fig.~\ref{fig:fig2}(g)-\ref{fig:fig2}(h).

\item[(5)]
The NLO twist-2 and twist-3 contributions to the form factors of $B \to \pi$
transitions have been completed very recently in Refs.\cite{prd85-074004,cheng14a}, the typical
Feynman diagrams are those as shown in Fig.~\ref{fig:fig2}(i)-\ref{fig:fig2}(l).
Based on the $SU(3)$ flavor symmetry, we could extend directly
the formulas for the NLO contributions to the form factor $F_{0,1}^{B \to \pi}(0)$
as given in Refs.~\cite{prd85-074004,cheng14a} to the cases for $B_s \to (K, \eta_s) $ transitions
after making some proper modifications for the relevant masses or decay constants of the mesons involved.

\item[(6)]
In Ref.~\cite{npb896-355}, we made the first calculation for the scalar pion form factors $F_{0,1}^{'(1)}$
up to the NLO level, which describes the LO and NLO ( ${\cal O} (\alpha_s^2)$ ) contributions to the
factorizable annihilation diagrams of the considered $B \to \pi\pi$ decays.
We found numerically that
(a) the NLO part of the form factor $F_{0,1}^{'(1)}$, i.e., the NLO annihilation correction,
is very small in size, but has a large strong phase
around $-55^0$,  and therefore may play an important role in producing large CP violation for
the relevant decay modes; and (b) the NLO annihilation correction can produce only a very small
enhancement (less than $3\%$ in magnitude) to their branching ratios for $B \to \pi^+\pi^-$ and
$\pi^0\pi^0$ decays ~\cite{npb896-355}.

\end{enumerate}

In this paper, we adopt directly the formulas for all currently known NLO contributions from
Refs.~\cite{npb675,nlo05,o8g2003,fan2013,prd85-074004,cheng14a,xiao14a,xiao14b,npb896-355}
without further discussions about the details.
At present, the calculations for the NLO corrections to the
LO hard spectator ( Fig.~\ref{fig:fig1}(c)-\ref{fig:fig1}(d) ) and the non-factorizable annihilation diagrams
( Fig.~\ref{fig:fig1}(g),\ref{fig:fig1}(h) ) have not been completed yet.
For most $B_s^0\to PP$ with $P=(\pi,K,\etap)$ as studied in Refs.~\cite{fan2013,xiao2014,xiao14a,xiao14b},
furthermore, we have made the comparative studies for the magnitude
of all relevant LO and NLO contributions from different kinds of Feynman diagrams
in great details and found that the LO contributions from the hard spectator
and annihilation diagrams are always much smaller than the corresponding dominant LO contribution
from the emission diagrams ( Fig.~\ref{fig:fig1}(a)-\ref{fig:fig1}(b) ),
those still unknown NLO contributions in the PQCD approach are in fact the higher order corrections
to the small LO pieces, and consequently should be much smaller than their LO counterparts in magnitude,
say less than $5\%$ of the dominant LO contribution, and could be neglected safely.
On the other hand, it is worth of mentioning that the uncertainty of current theoretical
predictions for the decay rates or CP violating asymmetries of those charmless hadronic two-body decays of $B/Bs$ mesons
is generally around $20$ to $30$ percent.

According to Refs.~\cite{npb675,nlo05,vc2006}, the vertex corrections can be absorbed
into the redefinition of the Wilson coefficients $a_i(\mu)$ by adding a vertex-function $V_i(M)$ to them.
\begin{eqnarray}
a_{1,2}(\mu) &\to& a_{1,2}(\mu) +\frac{\alpha_s(\mu)}{9\pi}\; C_{1,2}(\mu)\; V_{1,2}(M) \;, \non
a_i(\mu) &\to& a_i(\mu) +\frac{\alpha_s(\mu)}{9\pi}\; C_{i+ 1}(\mu) \; V_i(M), \qquad {\rm for\ \ } i=3,5,7,9, \non
a_j(\mu) &\to& a_j(\mu) +\frac{\alpha_s(\mu)}{9\pi}\; C_{j-1}(\mu) \; V_j(M), \qquad {\rm for \ \ } j=4,6,8,10,
\label{wnlo}
\end{eqnarray}
where $M$ denotes the meson emitted from the weak vertex ( i.e. the $M_2$ in Fig.~\ref{fig:fig2}(a)-\ref{fig:fig2}(d)).
For a pseudo-scalar meson $M$, the explicit expressions of the functions $V_i(M)$ have been
given in Eq.~(6) of Ref.~\cite{vc2006}. For the case of a vector meson $V$ one can obtain $V_i(V)$ from $V_i(P)$
by some appropriate replacements: $\phi^A \to \phi_V$, $\phi^P\to -\phi_V^s$ and the decay constant
$f_P \to f_V, f_V^T$ \cite{vc2006}.

The NLO ``Quark-Loop" and ``Magnetic-Penguin" contributions  are in fact a kind of penguin corrections
with the insertion of the four-quark operators and the chromo-magnetic operator $O_{8g}$ respectively,
as shown in Figs.~\ref{fig:fig2}(e,f) and \ref{fig:fig2}(g,h).
For the $b\to s$ transition, for example, the corresponding effective Hamiltonian $H_{eff}^{ql}$ and
$H_{eff}^{mp}$ can be written in the following form:
\beq
H_{eff}^{(ql)}&=&-\sum\limits_{q=u,c,t}\sum\limits_{q{\prime}}\frac{G_F}{\sqrt{2}}
V_{qb}^{*}V_{qs}\frac{\alpha_s(\mu)}{2\pi}C^{q}(\mu,l^2)\left[ \bar{b}\gamma_\rho
\left(1-\gamma_5\right)T^as\right ]\left(\bar{q}^{\prime}\gamma^\rho
T^a q^{\prime}\right),\label{eq:heff-ql}\\
H_{eff}^{mp} &=&-\frac{G_F}{\sqrt{2}} \frac{g_s}{8\pi^2}m_b\;
V_{tb}^*V_{ts}\; C_{8g}^{eff} \; \bar{s}_i \;\sigma^{\mu\nu}\; (1+\gamma_5)\;
 T^a_{ij}\; G^a_{\mu\nu}\;  b_j, \label{eq:heff-o8g}
\eeq
where $l^2$ is  the invariant mass of the gluon which attaches the quark loops
in Figs.~\ref{fig:fig2}(e,f), and the functions $C^{q}(\mu,l^2)$ can be found in Refs.~\cite{nlo05,xiao08b}.
The $C_{8g}^{eff}$ in Eq.~(\ref{eq:heff-o8g}) is the effective Wilson coefficient with
the definition of $C_{8g}^{eff}= C_{8g} + C_5$ \cite{nlo05}.

By analytical evaluations, we find the following two points:
\begin{enumerate}
\item[(1)] The four pure annihilation type decays $B_s^0 \to \pi\rho$ and $B_s^0 \to \pi\omega $
do not receive the NLO contributions from the vertex corrections, the quark-loop
and the magnetic-penguin diagrams. The only NLO contributions are included by using the NLO-WCs,
instead of the LO ones.

\item[(2)]
For the remaining seventeen decay channels, besides the LO decay amplitudes, one should take those NLO
contributions into account:
\begin{eqnarray}
{\renewcommand\arraystretch{2.0} \begin{array}{ll}
\displaystyle {\cal A}^{(u)}_{K^{0}\phi } \, \to\, {\cal
A}^{(u)}_{K^{0}\phi }+{\cal M}^{(u,c)}_{K\phi}\;,
&\displaystyle {\cal A}^{(t)}_{K^{0}\phi } \, \to\, {\cal
A}^{(t)}_{K^{0}\phi }-{\cal M}^{(t)}_{K\phi}-{\cal
M}^{(g)}_{K\phi}\;,
\\
\displaystyle {\cal A}^{(u)}_{ \rho^-K^{+}} \, \to\,
{\cal A}^{(u)}_{ \rho^-K^{+}}+{\cal M}^{(u,c)}_{\rho
K}\;, &\displaystyle {\cal A}^{(t)}_{ \rho^-K^{+}} \,
\to\, {\cal A}^{(t)}_{ \rho^-K^{+}}-{\cal
M}^{(t)}_{\rho K}-{\cal M}^{(g)}_{\rho K}\;,
\\
\displaystyle {\cal A}^{(u)}_{ \pi^-K^{*+}} \, \to\,
{\cal A}^{(u)}_{\pi^-K^{*+}}+{\cal M}^{(u,c)}_{\pi
K^*}\;, &\displaystyle {\cal A}^{(t)}_{\pi^-K^{*+}}
\, \to\, {\cal A}^{(t)}_{ \pi^-K^{*+}}-{\cal
M}^{(t)}_{\pi K^*}-{\cal M}^{(g)}_{\pi K^*}\;,
\\
\displaystyle {\cal A}^{(u)}_{\pi^0K^{*0} } \, \to\, {\cal
A}^{(u)}_{\pi^0K^{*0} } +\frac{1}{\sqrt{2}}{\cal M}^{(u,c)}_{\pi
K^*}\;, & \displaystyle {\cal A}^{(t)}_{\pi^0K^{*0} } \, \to\,
{\cal A}^{(t)}_{\pi^0K^{*0} } -\frac{1}{\sqrt{2}}{\cal
M}^{(t)}_{\pi K^*}-\frac{1}{\sqrt{2}}{\cal M}^{(g)}_{\pi K^*}\;,
\\
\displaystyle {\cal A}^{(u)}_{K^{\pm}K^{*\mp} } \, \to\,
{\cal A}^{(u)}_{K^{\pm}K^{*\mp} }+{\cal M}^{(u,c)}_{K^{\pm}K^{*\mp} }\;,
&\displaystyle {\cal A}^{(t)}_{K^{\pm}K^{*\mp}  } \, \to\,
{\cal A}^{(t)}_{K^{\pm}K^{*\mp}  }-{\cal M}^{(t)}_{K^{\pm}K^{*\mp} }-{\cal
M}^{(g)}_{K^{\pm}K^{*\mp} }\;,
\\
\displaystyle {\cal A}^{(u)}_{K^{*0}\eta_{s} } \, \to\,
{\cal A}^{(u)}_{K^{*0}\eta_{s} }+{\cal M}^{(u,c)}_{K^{*0}\eta_{s} }\;,
&\displaystyle {\cal A}^{(t)}_{K^{*0}\eta_{s}  } \, \to\,
{\cal A}^{(t)}_{K^{*0}\eta_{s}  }-{\cal M}^{(t)}_{K^{*0}\eta_{s} }-{\cal M}^{(g)}_{K^{*0}\eta_{s} }\;,
\\
\displaystyle {\cal A}^{(u)}_{\phi\eta_{s} } \, \to\,
{\cal A}^{(u)}_{\phi\eta_{s} }+{\cal M}^{(u,c)}_{\phi\eta_{s} }\;,
&\displaystyle {\cal A}^{(t)}_{\phi\eta_{s}  } \, \to\,
{\cal A}^{(t)}_{\phi\eta_{s}  }-{\cal M}^{(t)}_{\phi\eta_{s} }-{\cal M}^{(g)}_{\phi\eta_{s} }\;,
\\
\displaystyle {\cal A}^{(u)}_{\rho^0 K^{0} } \, \to\, {\cal A}^{(u)}_{\rho^0 K^{0}}
+\frac{1}{\sqrt{2}}{\cal M}^{(u,c)}_{\rho K}\;,& \displaystyle {\cal A}^{(t)}_{\rho^0 K^{0} } \, \to\,
{\cal A}^{(t)}_{\rho^0 K^{0}}
-\frac{1}{\sqrt{2}}{\cal M}^{(t)}_{\rho^0 K^{0}}-\frac{1}{\sqrt{2}}{\cal M}^{(g)}_{\rho^0 K^{0}}\;,
\\
\displaystyle {\cal A}^{(u)}_{ \omega K^{0} } \, \to\, {\cal A}^{(u)}_{ \omega K^{0}}
+\frac{1}{\sqrt{2}}{\cal M}^{(u,c)}_{\omega K}\;,& \displaystyle {\cal A}^{(t)}_{\omega K^{0} } \, \to\,
{\cal A}^{(t)}_{ \omega K^{0}}
-\frac{1}{\sqrt{2}}{\cal M}^{(t)}_{ \omega K^{0}}-\frac{1}{\sqrt{2}}{\cal M}^{(g)}_{ \omega K^{0}}\;,
\end{array} }
\end{eqnarray}
where the terms ${\cal A}_{M_2M_3}^{(u,t)}$ refer to the LO amplitudes,  while ${\cal M}_{M_2M_3}^{(u, c, t )}$
and ${\cal M}_{M_2M_3}^{(g)}$ are the NLO ones, which describe the NLO contributions from the up-loop, charm-loop,
QCD-penguin-loop, and magnetic-penguin diagrams, respectively.

\end{enumerate}

It is straightforward to calculate the decay amplitudes ${\cal M}_{M_2M_3}^{(ql)}$ and ${\cal M}_{M_2M_3}^{(mp)}$.
As mentioned in the previous section, since the Lorentz structure of wave functions for vector mesons
is different from those for pseudoscalar mesons, there are also two different kinds of decay amplitudes
${\cal M}_{M_2M_3}^{(ql)}$ and ${\cal M}_{M_2M_3}^{(mp)}$.
First, when the $M_2$ is a pseudoscalar meson and $M_3$ is a vector meson,
the NLO decay amplitudes  ${\cal M}_{PM_3}^{(ql)}$ and ${\cal M}_{PM_3}^{(mp)}$ can be written in the form:
\beq
{\cal M}^{(ql)}_{PV}&=& -8m_{B_s}^4\frac{{C_F}^2}{\sqrt{2N_c}} \int_0^1 dx_1dx_2dx_3
\int_0^\infty b_1db_1b_3db_3 \,\phi_{B_s}(x_1) \Bigl \{ \left [(1+x_3)\phi_p^A(x_2) \phi_v(x_3)
\right.
\non && \left. -2 r_p\phi_p^P(x_2) \phi_v(x_3)+ r_v(1-2x_3)\phi_p^A(x_2)(\phi_v^s(x_3)
+ \phi_v^t(x_3)) -2r_pr_v\phi_p^P(x_2)((2+x_3)\phi_v^s(x_3)\right.
\non &&
-x_3\phi_v^t(x_3))]\cdot \alpha_s^2(t_a) \cdot h_e(x_1,x_3,b_1,b_3)\cdot  \exp\left [-S_{ab}(t_a)\right ]\; C^{(q)}(t_a,l^2)+ [2r_v\phi_p^A(x_2)\phi_v^s(x_3)
\non &&
-4r_pr_v\phi_p^P(x_2)]\phi_v^s(x_3) \cdot
\alpha_s^2(t_b) \cdot h_e(x_3,x_1,b_3,b_1)
\cdot \exp[-S_{ab}(t_b)] \; C^{(q)}(t_b,l'^2)\Bigr \},
\eeq
\beq
{\cal M}^{(mp)}_{PV} &=& 16m_{B_s}^6\frac{{C_F}^2}{\sqrt{2N_c}} \int_0^1 dx_1dx_2dx_3
\int_0^\infty b_1db_1b_2db_2b_3db_3\, \phi_{B_s}(x_1)\non
&& \hspace{-1cm}\times \left \{ \left [-(1-x_3) \left [ 2\phi_v(x_3)- r_v(3\phi_v^s(x_3)
+\phi_v^T(x_3) )-r_v x_3(\phi_v^s(x_3)-\phi_v^T(x_3))\right ] \phi_p^A(x_2) \right.\right. \non
&& \hspace{-1cm} \left.\left.
- r_p x_2(1+x_3) (3\phi_p^P(x_2)-\phi_p^T(x_2))\phi_v(x_3)+r_pr_v(1-x_3)(3\phi_p^P(x_2)
+\phi_p^T(x_2))(\phi_v^s(x_3)-\phi_v^t(x_3))\right.\right.
\non &&\hspace{-1cm}
+r_pr_v x_2(1-2x_3)(3\phi_p^P(x_2)-\phi_p^T(x_2))(\phi_v^s(x_3)
+\phi_v^t(x_3))]\cdot \alpha_s^2(t_a) h_g(x_i,b_i)\cdot \exp[-S_{cd}(t_a)]\; C_{8g}^{eff}(t_a)
\non
&&  \hspace{-1cm} \left.
-[4r_v\phi_p^A(x_2)\phi_v^s(x_3)+2r_pr_v x_2(3\phi_p^P(x_2)-\phi_p^T(x_2))\phi_v^s(x_3) ]
\cdot \alpha_s^2(t_b) \cdot h'_g(x_i,b_i)\cdot \exp[-S_{cd}(t_b) ] \cdot C_{8g}^{eff}(t_b)\right \}.
 \eeq
When $M_2=V$ and $M_3=P$, however, the corresponding decay amplitudes can be written as
\beq
{\cal M}^{(ql)}_{VP}&=& -8m_{B_s}^4\frac{{C_F}^2}{\sqrt{2N_c}} \int_0^1 dx_1dx_2dx_3
\int_0^\infty b_1db_1b_3db_3 \,\phi_{B_s}(x_1)
\Bigl \{ \left [(1+x_3)\phi_v(x_2) \phi_p^A(x_3)
\right.
\non && \left.
+2 r_v\phi_v^s(x_2) \phi_P^A(x_3)+r_p(1-2x_3)\phi_v(x_2)(\phi_P^P(x_3)
+ \phi_p^T(x_3))+2r_pr_v\phi_v^s(x_2)((2+x_3)\phi_p^P(x_3) \right.
\non &&
-x_3\phi_P^T(x_3))]\cdot \alpha_s^2(t_a) \cdot h_e(x_1,x_3,b_1,b_3)\cdot  \exp\left [-S_{ab}(t_a)\right ]\; C^{(q)}(t_a,l^2)+ [2r_p\phi_v(x_2)\phi_p^P(x_3)
\non &&
+4r_pr_v\phi_v^s(x_2)]\phi_p^P(x_3) \cdot \alpha_s^2(t_b) \cdot h_e(x_3,x_1,b_3,b_1)
\cdot \exp[-S_{ab}(t_b)] \; C^{(q)}(t_b,l'^2)\Bigr \},
\eeq
\beq
{\cal M}^{(mp)}_{VP} &=& 16m_{B_s}^6\frac{{C_F}^2}{\sqrt{2N_c}} \int_0^1 dx_1dx_2dx_3
\int_0^\infty b_1db_1b_2db_2b_3db_3\, \phi_{B_s}(x_1)\non
&& \hspace{-1cm}\times \left \{ \left [-(1-x_3) \left [ 2\phi_p^A(x_3)+ r_p(3\phi_p^P(x_3)
+\phi_p^T(x_3) )+r_p x_3(\phi_p^P(x_3) -\phi_p^T(x_3))\right] \phi_v(x_2) \right.\right. \non
&& \hspace{-1cm} \left.\left.
- r_v x_2(1+x_3) (3\phi_v^s(x_2) -\phi_v^t(x_2))\phi_p^A(x_3)-r_pr_v(1-x_3)(3\phi_v^s(x_2)
+\phi_v^t(x_2))(\phi_p^P(x_3) -\phi_p^T(x_3)) \right. \right.
\non &&\hspace{-1cm}
-r_pr_v x_2(1-2x_3)(3\phi_v^s(x_2)-\phi_v^t(x_2))(\phi_p^P(x_3)
+\phi_p^T(x_3))]\cdot \alpha_s^2(t_a) h_g(x_i,b_i)\cdot \exp[-S_{cd}(t_a)]\; C_{8g}^{eff}(t_a)
\non
&&  \hspace{-1cm} \left.
- [4r_p\phi_v(x_2)\phi_p^P(x_3)+2r_pr_v x_2(3\phi_v^s(x_2)-\phi_v^T(x_2))\phi_p^P(x_3)]
\cdot \alpha_s^2(t_b) \cdot h'_g(x_i,b_i)
\cdot \exp[-S_{cd}(t_b)]\; \cdot C_{8g}^{eff}(t_b)\right \} .
\eeq
The explicit expressions for the hard functions $(h_e,h_g,h'_g)$, the functions
$C^{(q)}(t_a,l^2)$ and $C^{(q)}(t_b,l'^2)$, the Sudakov functions $S_{ab}(t)$ and $S_{cd}(t)$,
the hard scales $t_{a,b}$ and the effective Wilson coefficients $C_{8g}^{eff}(t)$,
can be found easily for example in Refs.~\cite{nlo05,fan2013,xiao14a,xiao14b}.

As mentioned in previous section, the NLO twist-2 and twist-3 contributions to the form factors of
$B \to \pi$ transition have been calculated very recently in Refs.~\cite{prd85-074004,cheng14a}.
Based on the approximation of the $SU(3)$ flavor symmetry, we extend the formulas for $B\to \pi$
transitions as given in Refs.~\cite{prd85-074004,cheng14a} to the cases for
$B_s \to (K, \eta_s)$ transition form factors directly, after making appropriate
replacements  for some relevant parameters.
The NLO form factor $f^+(q^2)$ for $B_s \to K$ transition, for example, can be written in the form:
\beq
f^+(q^2)|_{\rm NLO} &=& 8 \pi m^2_{B_s} C_F \int{dx_1 dx_2} \int{b_1 db_1 b_2 db_2}
\phi_{B_s}(x_1,b_1)\non &&
\hspace{-1cm}\times \Biggl \{ r_K \left [\phi_{K}^{P}(x_2) - \phi_{K}^{T}(x_2) \right ]
\cdot \alpha_s(t_1)\cdot e^{-S_{B_s K}(t_1)}\cdot S_t(x_2)\cdot h(x_1,x_2,b_1,b_2) \non
&&\hspace{-1cm}  + \Bigl [ (1 + x_2 \eta)
\left (1 + F^{(1)}_{\rm T2}(x_i,\mu,\mu_f,q^2)\; \right ) \phi_{K}^A(x_2)
+ 2 r_K \left (\frac{1}{\eta} - x_2 \right )\phi_{K}^T(x_2)
- 2x_2 r_K \phi_{K}^P(x_2) \Bigr ] \non
&& \hspace{-1cm} \cdot \alpha_s(t_1)\cdot e^{-S_{B_s K}(t_1)} \cdot S_t(x_2)\cdot h(x_1,x_2,b_1,b_2)\non
&& \hspace{-1cm} + 2 r_{K} \phi_{K}^P(x_2) \left (1 + F^{(1)}_{\rm T3}(x_i,\mu,\mu_f,q^2)\right )
\cdot \alpha_s(t_2)\cdot e^{-S_{B_s K}(t_2)} \cdot S_t(x_2)\cdot h(x_2,x_1,b_2,b_1) \Biggr \},
\label{eq:ffnlop}
\eeq
where $\eta=1-q^2/m_{B_s}^2$ with $q^2=(P_{B_s}-P_3)^2$ and $P_3$ is the momentum of the
meson $M_3$ which absorbed the spectator $\bar{s}$ quark of the $\bar{B}^0_s$ meson, $\mu$ ($\mu_f$) is  the
renormalization (factorization ) scale, the hard scale $t_{1,2}$ are
chosen as the largest scale of the propagators in the hard $b$-quark decay diagrams
\cite{prd85-074004,cheng14a}. The explicit expressions of the threshold Sudakov function $S_t(x)$ and the hard function
$h(x_i,b_j)$ can be found in Refs.~\cite{prd85-074004,cheng14a}.
The NLO correction factor $F^{(1)}_{\rm T2}(x_i,\mu,\mu_f,q^2)$ and
$F^{(1)}_{\rm T3}(x_i,\mu,\mu_f,q^2)$ appeared in Eq.~(\ref{eq:ffnlop})
describe the NLO twist-2 and twist-3 contributions to the form factor $f^{+,0}(q^2)$ of the
$B_s \to K$ transition respectively,  and can be written in the following form \cite{prd85-074004,cheng14a}:
\beq
F^{(1)}_{\rm T2}&=& \frac{\alpha_s(\mu_f) C_F}{4 \pi}
\Biggl [\frac{21}{4} \ln{\frac{\mu^2}{m^2_{B_s}}}
-(\frac{13}{2} + \ln{r_1}) \ln{\frac{\mu^2_f}{m^2_{B_s}}}
+\frac{7}{16} \ln^2{(x_1 x_2)}+ \frac{1}{8} \ln^2{x_1} \non
&&+ \frac{1}{4} \ln{x_1} \ln{x_2}
+ \left (- \frac{1}{4}+ 2 \ln{r_1} + \frac{7}{8} \ln{\eta} \right ) \ln{x_1}
+ \left (- \frac{3}{2} + \frac{7}{8} \ln{\eta} \right) \ln{x_2} \non
&&+ \frac{15}{4} \ln{\eta} - \frac{7}{16} \ln^{2}{\eta}
+ \frac{3}{2} \ln^2{r_1} - \ln{r_1}
+ \frac{101 \pi^2}{48} + \frac{219}{16} \Biggr ],  \label{eq:ffnlot2}\\
F^{(1)}_{\rm T3}&=&\frac{\alpha_s(\mu_f) C_F}{4 \pi}
\Biggl [\frac{21}{4} \ln{\frac{\mu^2}{m^2_{B_s}}}
- \frac{1}{2}(6 + \ln{r_1}) \ln{\frac{\mu^2_f}{m^2_{B_s}}}
+ \frac{7}{16} \ln^2{x_1} - \frac{3}{8} \ln^2{x_2} \non
&& + \frac{9}{8} \ln{x_1} \ln{x_2}
+ \left (- \frac{29}{8}+ \ln{r_1} + \frac{15}{8} \ln{\eta} \right ) \ln{x_1}
+ \left (- \frac{25}{16} + \ln{r_2} + \frac{9}{8} \ln{\eta} \right) \ln{x_2} \non
&& + \frac{1}{2} \ln{r_1} - \frac{1}{4} \ln^{2}{r_1} + \ln{r_2}
- \frac{9}{8} \ln{\eta} - \frac{1}{8} \ln^{2}{\eta} + \frac{37 \pi^2}{32}
+ \frac{91}{32} \Biggr ],
\eeq
where $r_i=m^2_{B_s}/\xi_i^2$ with the choice of $\xi_1=25 m_{B_s}$
and $\xi_2=m_{B_s}$. For the $B_s \to (K, \eta_s)V$ decays,
the large recoil region corresponds to the energy fraction  $\eta  \sim \textit{O}(1)$.
The factorization scale $\mu_f$ is set to be the hard scales
\beq
t^{a}=\max(\sqrt{x_3 \eta } \, m_B ,1/b_1,1/b_3), \quad or \quad
t^{b}=\max(\sqrt{x_1 \eta } \, m_B ,1/b_1,1/b_3),
\eeq
corresponding to the largest energy scales in Fig.~\ref{fig:fig1}(a)
and \ref{fig:fig1}(b), respectively. The renormalization scale $\mu$ is defined as \cite{prd83-054029,fan2013,cheng14a}
\begin{eqnarray}
\mu = t_s (\mu_{\rm f})  = \left \{{\rm Exp} \left[ c_1 + \left(\ln
{m_B^2 \over \zeta_1^2}  +{5 \over 4} \right)  \ln{\mu_{\rm f}^2
\over m_B^2 } \right ]  \, x_1^{c_2 } \, x_3^{c_3} \right \}^{2/21}
\, \mu_{\rm f}, \label{ts function}
\end{eqnarray}
with the coefficients
\begin{eqnarray}
c_1 &=& - \left ({15 \over 4} -{7 \over 16} \ln \eta \right ) \ln
\eta + {1 \over 2} \ln { m_B^2 \over\zeta_1^2 }   \left ( 3 \ln {
m_B^2 \over\zeta_1^2 } + 2 \right ) - {101 \over 48} \pi^2 - {219 \over 16} \,,  \non
c_2 &=& - \left ( 2 \ln { m_B^2 \over\zeta_1^2 }
+ {7 \over 8} \ln \eta - {1 \over 4} \right ) ,\non
c_3 &=& -{7 \over 8} \ln \eta + {3 \over 2}.
\end{eqnarray}

\section{Numerical results}\label{sec:n-d}

In the numerical calculations, the following input parameters will be used implicitly.
The masses, decay constants and QCD scales are in units of GeV \cite{pdg2016}:
\beq
\Lambda_{\overline{\mathrm{MS}}}^{(f=5)} &=& 0.225,
\quad  f_{B_{s}} = 0.23\pm 0.02, \quad  f_K = 0.16, \quad   f_{\pi} = 0.13, \quad
f_{\rho}=0.209,\quad f_{\rho}^{T}=0.165, \non
M_{B_{s}} &=&  5.37,\quad   m_K=0.494,\quad m_0^\pi = 1.4, \quad   m_0^K = 1.9, \quad
f_{\omega}=0.195, \quad f_{\omega}^{T}=0.145,\non
f_{K^*}&=&0.217,\quad f_{K^*}^{T}=0.185,\quad f_{\phi}=0.231, \quad f_{\phi}^{T}=0.20, \quad
m_{\rho}= 0.77,\quad m_{\omega}=0.78,\non
\quad m_{K^*}&=& 0.89,\quad m_{\phi}=1.02, \quad
\tau_{B_s^0} = 1.497 {\rm ps}, \quad m_b=4.8, \quad M_W = 80.42. \label{eq:para}
\eeq
For the CKM matrix elements, we also take the same values
as being used in Ref.~\cite{ali07}, and neglect the small errors on
$V_{ud}, V_{us}$, $V_{ts}$ and $V_{tb}$
\beq
|V_{ud}|&=& 0.974, \quad |V_{us}|=0.226,
\quad |V_{ub}|=\left ( 3.68^{+0.11}_{-0.08}\right)\times 10^{-3},\quad
|V_{td}|=\left ( 8.20^{+0.59}_{-0.27}\right)\times 10^{-3},\non
|V_{ts}|&=& 40.96\times 10^{-3}, \quad |V_{tb}|= 1.0, \quad
\alpha = (99^{+4}_{-9.4})^\circ ,\quad \gamma=(59.0^{+9.7}_{-3.7})^\circ.
\label{eq:angles}
\eeq

For the considered $B_s^0$ decays, the decay amplitude for a given decay mode
with $b \to q$ transitions can be generally written as
\beq
{\cal A}(\bar{B}_s^0 \to f) &=& V_{ub}V_{uq}^* T -V_{tb}V_{tq}^*  P
= V_{ub}V_{uq}^*T  \left [ 1 + z e^{ i ( -\theta + \delta ) } \right],\label{eq:ma}
\eeq
where $q=(d,s)$, $\theta$ is the weak phase ( the CKM angles ),
$\delta=\arg[P/T]$  are the relative strong phase between the tree ($T$)
and penguin ($P$) diagrams, and the parameter  ``$z$"  is the ratio of penguin to
tree contributions with the definition
\beq
z=\left|\frac{V_{tb} V_{tq}^*}{ V_{ub}V_{uq} ^*} \right| \left|\frac{P}{T}\right|,
\eeq
the ratio $z$ and the strong phase $\delta$ can be calculated in the
PQCD approach. Therefore the CP-averaged branching ratio, consequently, can be
defined as
\beq
{\cal B}(\bar{B}_s^0\to f) \propto \frac{1}{2} \left [ |{\cal A}|^2 +|{\overline{\cal A}}|^2\right]
=| V_{ub}V_{uq} ^*T|^2 \left [1+2z\cos\theta\cos\delta+z^2\right],
\label{eq:br0}
\eeq
where the ratio $z$ and the strong phase $\delta$ have been defined in the above equations.

In Table \ref{tab:br1}, we list the PQCD predictions for the CP-averaged branching ratios
of the considered $B_s^0$ decays. The label``LO" means the full leading order PQCD predictions.
For other four cases with the label ``$+$VC" , ``$+$QL",  ``$+$MP" and ``NLO" , the NLO Wilson coefficients
$C_i(\mu)$ and $\alpha_s(\mu)$ at two-loop level are used implicitly.
The label ``$+$VC" means the additional NLO "Vertex correction" is included.
The label ``$+$QL" ("$+$MP") means both  "VC" and "QL" (  "VC" , "QL" and "MP" ) NLO contributions
are taken into account simultaneously.
And finally the label ``NLO" means that all currently known NLO contributions  are taken into account:
the newly known NLO corrections to the form factor $F_0^{B_s\to K}(0) $ and $F_0^{B_s\to \eta_s}(0)$ also be
included here.
In Table \ref{tab:br1}, for the sake of comparison, we also list the LO PQCD predictions (in the seventh column)
as given in Ref.~\cite{ali07}, the QCDF predictions as given in
Ref.~\cite{npb675} (the eighth column ) and in Ref.~\cite{chengbs09} ( the ninth column )  respectively.
The main theoretical errors come from the uncertainties of the various input parameters: dominant ones
from $\omega_{B_s}=0.50 \pm 0.05$ GeV, $f_{B_s}=0.23\pm 0.02$ GeV and the Gegenbauer moments like $a_2^{\pi,k}=0.25\pm 0.15$.
The total errors of the NLO PQCD predictions as listed in Table \ref{tab:br1} are obtained
by adding the individual errors in quadrature.

Among the twenty one $B_s^0 \to PV$ decays considered in this paper, only three of them,
say $\bar{B}^0_s \to \pi^- K^{*+}, K^+ K^{*-} $ and $\bar{B}^0_s \to K^0 \bar{K}^{*0} $,
have been measured recently by LHCb experiments \cite{lhcb-d1}.
For $\bar{B}_s^0 \to \etar \phi$ decay, the LHCb Collaboration put an upper limit at $95\%$ C.L.
on its decay rate very recently \cite{lhcb-d2}.
We list those measured values and upper limit in the last column of  Table \ref{tab:br1}
and will compare those theoretical predictions with them.

\begin{table}
\caption{ The PQCD predictions for the CP-averaged branching ratios ( in units of $10^{-6}$ ) of the considered
$\bar{B}_s^0$ decays. As a comparison, we also list the theoretical
predictions as given in Refs.~\cite{ali07,npb675,chengbs09}, and those currently available measured values
\cite{lhcb-d1} or upper limit at $95\%$ C.L. \cite{lhcb-d2}.}
\label{tab:br1}
\begin{tabular*}{16cm}{@{\extracolsep{\fill}}l|l|lllll|lll|l} \hline\hline
{\rm Mode} &{\rm Class}&{\rm LO} &{\rm + VC}& {\rm + QL}&{\rm + MP} & {\rm NLO}& {\rm PQCD}\cite{ali07}
& {\rm QCDF 1 }\cite{npb675}& {\rm QCDF 2 }\cite{chengbs09} & {\rm Data}\cite{lhcb-d1,lhcb-d2}
 \\ \hline
 $\bar B_s^0\to \pi^-K^{*+}$&      {\rm T}  & $6.32    $&$5.12  $&$4.01  $&$3.96        $&$3.96^{+1.41}_{-1.16}$&$7.6^{+3.0}_{-2.3}$&$8.7^{+5.8}_{-4.9}$
 & $7.8^{+0.6}_{-1.0}$ &$3.3\pm 1.2 $\\
 $\bar B_s^0\to K^+K^{*-} $& {\rm P}  & $9.03    $&$10.75  $&$12.30 $&$12.24     $&$12.23^{+2.95}_{-3.41}$&$10.7^{+5.2}_{-3.5}$
 &$9.6^{+24.7}_{-7.9} $&$10.3^{+5.7}_{-4.7} $&$12.5\pm 2.6 $\\
 $\bar B_s^0\to K^{0}\bar K^{*0} $& {\rm P}  & $9.95  $&$12.41 $&$14.46 $&$14.38  $&$14.39^{+3.54}_{-2.93}$&$11.6^{+5.5}_{-3.6}$
 &$8.1^{+24.6}_{-7.5}$& $ 10.5^{+6.1}_{-5.3}$ &$16.4\pm 4.1$\\ \hline
$\bar B_s^0\to K^+ \rho^- $&     {\rm T}&  $18.6  $&$16.3  $&$16.6  $&$16.4     $&$15.9^{+6.5}_{-4.9}        $&$17.8^{+7.89}_{-5.89}   $
&$24.5^{+15.2}_{-12.9}$ & $14.7^{+1.7}_{-2.3}$&$-$\\
$\bar B_s^0\to {\pi}^{0}K^{*0}$& {\rm C}&  $0.08   $&$0.20  $&$0.20  $&$0.21    $&$0.21^{+0.07}_{-0.04}        $&$0.07^{+0.04}_{-0.02}   $&$0.25^{+0.46}_{-0.22}$
& $0.89^{+1.16}_{-0.49}$ &$-$\\
$\bar B_s^0\to \eta {\phi}$&     {\rm P}&  $3.3  $&$0.89  $&$1.40 $&$1.21      $&$1.26^{+0.31}_{-0.23}   $&$3.6^{+1.7}_{-1.2}    $&$0.12^{+1.13}_{-0.26}$
& $1.0^{+1.6}_{-1.2}$ &$-$\\
$\bar B_s^0\to \etar {\phi} $&   {\rm P}&  $0.25   $&$0.63    $&$0.76 $&$0.51   $&$0.59^{+0.10}_{-0.13}         $&$0.19^{+0.20}_{-0.13} $&$0.05^{+1.18}_{-0.19}$
& $2.2^{+9.4}_{-2.2}$ &$< 1.01$\\
$\bar B_s^0\to K^0\phi$&         {\rm P}&  $0.20   $&$0.22  $&$0.23  $&$0.23    $&$0.24\pm 0.05        $&$0.16^{+0.10}_{-0.05} $&$0.27^{+0.74}_{-0.25} $
& $0.6^{+0.7}_{-0.4}$ &$-$\\ \hline
$\bar B_s^0\to K^0 {\rho}^{0}$&  {\rm C}&  $0.10   $&$0.39  $&$0.36  $&$0.34        $&$0.34^{+0.12}_{-0.09}        $&$0.08^{+0.07}_{-0.04}   $&$0.61^{+1.26}_{-0.61}$
& $1.9^{+3.2}_{-1.1}$ &$-$ \\
$\bar B_s^0\to K^{0}\omega$&     {\rm C}&  $0.14   $&$0.51  $&$0.56  $&$0.62         $&$0.65^{+0.22}_{-0.17}        $&$0.15^{+0.08}_{-0.05}   $&$0.51^{+0.83}_{-0.40}$
& $1.6^{+2.4}_{-0.9}$ &$-$ \\
$\bar B_s^0\to \eta K^{*0}$&    {\rm C} &  $0.20  $&$0.22  $&$0.22  $&$0.22    $&$0.20^{+0.04}_{-0.03}           $&$0.17^{+0.11}_{-0.07}    $&$0.26^{+0.78}_{-0.30}$
& $0.56^{+0.48}_{-0.22}$ &$-$\\
$\bar B_s^0\to \etar K^{*0}$&   {\rm C} & $0.11   $&$0.26 $&$0.33  $&$0.37       $&$0.35^{+0.05}_{-0.03} $&$0.09^{+0.04}_{-0.03}   $&$0.28^{+0.59}_{-0.30}$
& $0.90^{+1.00}_{-0.51}$ &$-$\\ \hline
$\bar B_s^0\to \pi^0\phi$  &    {\rm P$_{\rm EW}$} &  $0.13   $&$0.11  $&$-     $&$-            $&$0.11^{+0.05}_{-0.02}        $&$0.12^{+0.06}_{-0.05}   $&$0.16^{+0.05}_{-0.05}$
& $0.12^{+0.05}_{-0.02}$ &$-$\\
$\bar B_s^0\to \eta {\rho}^{0}$&  {\rm  P$_{\rm EW}$}  & $0.08   $&$0.12  $&$-  $&$-         $&$0.11^{+0.02}_{-0.02}     $&$0.06^{+0.03}_{-0.02}   $&$0.17^{+0.08}_{-0.07}$
& $0.10^{+0.03}_{-0.02}$ &$-$\\
$\bar B_s^0\to \etar {\rho}^{0}$& {\rm  P$_{\rm EW}$}  & $0.13   $&$0.20  $&$- $&$-         $&$0.19^{+0.05}_{-0.03}        $&$0.13^{+0.06}_{-0.04}   $&$0.25^{+0.12}_{-0.09}$
& $0.16^{+0.07}_{-0.04}$ &$-$ \\
$\bar B_s^0\to \eta {\omega}$&    {\rm P, C}  & $0.07   $&$0.11  $&$-     $&$-            $&$0.11^{+0.04}_{-0.03}        $&$0.04^{+0.06}_{-0.02} $&$0.012^{+0.030}_{-0.009}$
& $0.03^{+0.13}_{-0.02}$&$-$ \\
$\bar B_s^0\to \etar {\omega}$&   {\rm P, C}  & $0.30   $&$0.35  $&$- $&$-         $&$0.35^{+0.06}_{-0.04}  $&$0.44^{+0.23}_{-0.19} $&$0.024^{+0.092}_{-0.021} $
& $0.15^{+0.31}_{-0.10}$ &$-$\\ \hline
$\bar B_s^0\to \pi^0 \omega$&   {\rm ann} &  $0.004  $&$-  $&$-  $&$-             $&$0.003     $&$0.004   $&$0.0005 $
&$-$&$-$\\
$\bar B_s^0\to \pi^- \rho^+$&   {\rm ann} &  $0.22   $&$-    $&$-    $&$-             $&$0.13^{+0.04}_{-0.03}    $&$0.22^{+0.06}_{-0.07} $&$0.003$
& $0.02\pm 0.01$ &$-$\\
$\bar B_s^0\to \pi^+\rho^-$&    {\rm ann} &  $0.26   $&$-    $&$-    $&$-             $&$0.12^{+0.04}_{-0.03}    $&$0.24^{+0.07}_{-0.07}   $&$0.003 $
& $0.02\pm 0.01$&$-$\\
$\bar B_s^0\to \pi^0\rho^0$&    {\rm ann} &  $0.24    $&$-    $&$-    $&$-              $&$0.12^{+0.04}_{-0.03}  $&$0.23^{+0.07}_{-0.08}    $&$0.003  $
& $0.02\pm 0.01$ &$-$\\
\hline\hline
\end{tabular*}
\end{table}

From our PQCD predictions for the branching ratios, the previous theoretical predictions as given in
Refs.~\cite{ali07,npb675,chengbs09}
and the  data \cite{lhcb-d1,lhcb-d2}, as listed in Table \ref{tab:br1}, we have the following observations:
\begin{itemize}
\item[(1)]
The LO PQCD predictions for branching ratios of $\bar{B}^0_s \to (\pi,K,\etap) V$ decays as given
in Ref.~\cite{ali07} ten years ago  are confirmed by our independent calculations.
Some little differences between the central values of the LO predictions
are induced by the different choices or upgrade of some input parameters, such as the Gagenbauer moments
and the CKM matrix elements.

\item[(2)]
For the ``QCD-Penguin" decays $\bar B_s^0  \to K^0 \bar{K}^{*0}$  and $\bar B_s^0  \to K^{\pm}K^{*\mp}$, the NLO
contributions can provide $\sim 30\%$ to $ 45\%$ enhancements to the LO PQCD predictions of their
branching ratios. For the ``tree" dominated decay $\bar B_s^0\to \pi^{-} K^{*+}$, however,
the NLO contribution will result in a $37\%$ reduction of the LO PQCD prediction for its branching ratio.
The resultant enhancements or the reduction, fortunately, are all in the right direction.
After the inclusion of the NLO corrections, the NLO PQCD predictions for these three decays
become well consistent with those currently available data within one standard deviation.
In order to show numerically the improvements due to inclusion of the  NLO corrections,
we define the ratios $R_{1,2,3}$ of the measured values and the PQCD predictions for those  three measured decay modes:
\beq
R_1&=&\frac{{\cal B}( \bar{B}_s^0\to \pi^-K^{*+})^{\rm exp}}{ {\cal B}( \bar{B}_s^0\to \pi^-K^{*+})^{\rm PQCD}}
\approx \left\{\begin{array}{ll} 0.52,  & {\rm LO},\\ 0.83,  & {\rm NLO},\\  \end{array} \right. \\
R_2&=&\frac{{\cal B}( \bar{B}_s^0\to K^+ K^{*-})^{\rm exp}}{ {\cal B}( \bar{B}_s^0\to K^+K^{*-})^{\rm PQCD}}
\approx \left\{\begin{array}{ll} 1.38,  & {\rm LO},\\ 1.02,  & {\rm NLO},\\
\end{array} \right. \\
R_3&=&\frac{{\cal B}( \bar{B}_s^0\to K^0 \bar{K}^{*0})^{\rm exp}}{ {\cal B}(
\bar{B}_s^0\to K^0 \bar{K}^{*0}  )^{\rm PQCD}}
\approx \left\{\begin{array}{ll} 1.65,  & {\rm LO},\\ 1.14,  & {\rm NLO}.\\
\end{array} \right.
\label{eq:r123}
\eeq
It is easy to see that the agreements between the PQCD predictions and the three measured values
are indeed improved significantly due to the inclusion of the NLO contributions.
This is a clear indication for the
important role of the NLO contributions in order to understand the experimental measurements.

\item[(3)]
For the ``tree" dominated decay $\bar B_s^0\to K^+ \rho^-$, the NLO contribution results in
a $\sim 15\%$ reduction against the LO result, but its branching ratio is still at $1.6\times 10^{-5}$ level,
the largest one of all decays considered in this paper. We believe that this decay mode could be measured
by LHCb soon. For the ``color-suppressed-tree" decay
$\bar B_s^0\to \pi^0 K^{*0}$, however, although the NLO contribution can provide a large
$\sim 150\%$ enhancement, but the theoretical predictions for
its branching ratio in the LO and NLO PQCD or in the QCDF approaches\cite{npb675,chengbs09} are always at the
level of $ 10^{-7}$, much smaller than that for $\bar B_s^0\to K^+ \rho^-$ decay.
At the LO and NLO level, one can read out the ratio of the branching ratios of these two decays from Table
\ref{tab:br1}
\beq
R_4&=&\frac{{\cal B}( \bar{B}_s^0\to K^+ \rho^-) }{ {\cal B}(\bar{B}_s^0\to \pi^0 K^{*0}  ) }
\approx \left\{\begin{array}{ll} 232,  & {\rm LO},\\ 76,  & {\rm NLO}.\\
\end{array} \right. \label{eq:R4br}
\eeq
In order to understand so large difference, we made careful examinations for the LO decay amplitudes
of these two decay modes and found the two reasons.
Firstly, as shown explicitly in Eqs.~(\ref{eq:a002},\ref{eq:a003}), the dominant part of the decay amplitudes
for these two decays are very different in magnitude (in units of $10^{-4}$):
\beq
{\cal A}_T(\bar B_s^0\to K^+ \rho^-)&=& V_{ub}V_{ud}^{*} \left [ f_{\rho} F_{e K}\; a_1 + M_{e K} \; C_{1}
\right ] =17.45 -49.38\; i\ \ , \\
{\cal A}_C(\bar B_s^0\to \pi^0 K^{*0})&=& V_{ub}V_{ud}^{*} \left [ f_{\pi} F_{eK^*} \; a_{2} + M_{e K^*}\; C_{2}
\right ]/\sqrt{2} =-2.15 +0.42 \; i\ \ ,
\eeq
where $a_1=C2+C_1/3\approx C_2 \approx 1.1$ is a large quantity, while $|a_2|\approx |C_1 +C_2/3| \sim 0.1$ a small one.
The ratio of these two magnitudes $|{\cal A}_T|/|\cala_C|\approx 33$ is therefore very large. This is the main reason
of the large difference of these two branching ratios.
Secondly, there is a strong constructive interference
among the large ${\cal A}_T$ and ${\cal A}_P$ for $\bar B_s^0\to K^+ \rho^-$, but
a destructive one between the small ${\cal A}_C$ and ${\cal A}_P$ for $\bar B_s^0\to \pi^0 K^{*0}$ decay.
Numerically, one finds ( in units of $10^{-4}$) that
\beq
{\cal A}(\bar B_s^0\to K^+ \rho^-)^{\rm LO}&=& \underbrace{(17.45 -49.38\; i)}_{{\cal A}_T}
+ \underbrace{(6.39 - 2.00 \; i)}_{{\cal A}_P} = 23.83-51.38\; i, \\
{\cal A}(\bar B_s^0\to \pi^0 K^{*0})^{\rm LO}&=&
\underbrace{(-2.15 +0.42 \; i)}_{{\cal A}_C} + \underbrace{(1.19 - 2.34 \; i)}_{{\cal A}_P}
= -0.96-1.92 \; i.
\eeq
For the corresponding CP-conjugated decay modes, we also find similar behaviour
\beq
{\cal A}(B_s^0\to K^- \rho^+)^{\rm LO}&=& \underbrace{(15.5 +50.0\; i)}_{{\cal A}_T}
+ \underbrace{(3.24 - 5.84 \; i)}_{{\cal A}_P} = 18.7+44.2\; i, \\
{\cal A}(B_s^0\to \pi^0 \bar{K}^{*0})^{\rm LO}&=&
\underbrace{(1.473 -1.62 \; i)}_{{\cal A}_C} + \underbrace{(-0.75 - 2.51 \; i)}_{{\cal A}_P} = 0.72-4.13 \; i.
\eeq
From above four decay amplitudes, it is simple to define the ratio $R_4(|\cala|^2)$ of the square of
the decay amplitudes:
\beq
R_4(|{\cal A}|^2)^{\rm LO}=\frac{  |{\cal A}(\bar B_s^0\to K^+ \rho^-)^{\rm LO}|^2
+  |{\cal A}(B_s^0\to K^- \rho^+)^{\rm LO}|^2}{
| {\cal A}(\bar B_s^0\to \pi^0 K^{*0})^{\rm LO} |^2 + | {\cal A}(B_s^0\to \pi^0 \bar{K}^{*0})^{\rm LO} |^2 }
\approx 248,
\eeq
which is indeed close to the ratio of the CP-averaged branching ratios: $R_4^{\rm LO} \approx 232$ as
defined in Eq.~(\ref{eq:R4br}).
From above numerical results, it is straightforward to understand the large difference
between the LO PQCD predictions for ${\cal B}(\bar B_s^0\to K^+ \rho^-)$ and
${\cal B}(\bar B_s^0\to \pi^0 K^{*0})$. At the NLO level, the ratio $R_4^{\rm NLO}\approx 76$
can be interpreted in a similar way.

\item[(4)]
The two $\bar B_s^0  \to \phi \eta, \phi \etar $  decays are very similar in nature,
the difference between the PQCD predictions for ${\cal B}(\bar B_s^0  \to \phi \eta)$ and  ${\cal B}(\bar B_s^0  \to \phi
\etar)$ is rather large at LO level: $R_5^{\rm LO}(\calb)=3.3/0.25\approx 13.2$, but become smaller at NLO level:
$R_5^{\rm NLO}(\calb)=1.26/0.59\approx 2.14$, after the inclusion of the NLO contributions.
For $\bar B_s^0  \to \phi \eta $ decays, the NLO contribution results in a $62\%$ reduction for its branching ratio.
For $\bar B_s^0  \to \phi \etar$ decay, however, the inclusion of the NLO contribution leads to a $136\%$
enhancement to its LO result.
How to understand these special features for these two decay modes? The major reason is the unique
$\eta-\etar$ mixing pattern.
We know that the decay amplitude for $\bar{B}_s^0\to V (\eta, \etar)$ with $V=(\phi, \omega, \rho^0, K^{*0})$
can be written as
\beq
{\cal A}(\bar B_s^0  \to V \eta ) &=& {\cal A}( V\eta_q ) \cos(\phi) - {\cal A}(V \eta_s) \sin(\phi) , \label{eq:ee1}\\
{\cal A}(\bar B_s^0  \to V \etar ) &=& {\cal A}(V \eta_q ) \sin(\phi) + {\cal A}(V\eta_s) \cos(\phi), \label{eq:ee2}
\eeq
where $\phi=39.3^0$ is the mixing angle of $\eta-\etar$ system \cite{fks98}.
Since $\sin(\phi)=0.63$ has the same sign with $\cos(\phi)=0.77$ and are similar in
magnitude, the interference between the two parts, consequently,  may be constructive for
one channel but destructive for another, or vise versa.

For  $\bar B_s^0  \to \phi \eta$ and $\phi \etar $ decays, for example, we find the LO PQCD predictions
for their decay amplitudes (in units of $10^{-4}$)
\beq
{\cal A}(\bar B_s^0  \to \phi \eta )^{\rm LO} &=& \underbrace{(13.25+4.23\; i)}_{{\cal A}( \phi\eta_q )} \cdot \cos(39.3^\circ)
-\underbrace{(-17.16-8.61\; i)}_{{\cal A}(\phi \eta_s)} \cdot\sin(39.3^\circ)
= 21.01 +8.68\; i, \label{eq:ee11}\\
{\cal A}(\bar B_s^0  \to \phi \etar )^{\rm LO} &=& \underbrace{(13.25+4.23\; i)}_{{\cal A}( \phi\eta_q )} \cdot \sin(39.3^\circ)
+\underbrace{(-17.16-8.61\; i)}_{{\cal A}(\phi \eta_s)} \cdot\cos(39.3^\circ)
= -4.86 -3.96\; i.  \label{eq:ee12}
\eeq
And it is easy to see that $\cala(\phi \eta_q)$ interfere constructively
with $\cala(\phi \eta_s)$ for $\bar B_s^0  \to \phi \eta$ decay, but destructively with $\cala(\phi \eta_s)$
for $\bar B_s^0  \to \phi \etar$ decay. Such pattern of interference leads to the large ratio of $|\cala|^2$
\beq
R_5^{\rm LO}(|\cala |^2)= \frac{|{\cal A}(\bar B_s^0  \to \phi \eta )^{\rm LO}|^2 }{
|{\cal A}(\bar B_s^0  \to \phi \etar )^{\rm LO}|^2}=13.14\approx R_5^{LO}(\calb).
\eeq
When the NLO contributions are taken into account, however, we find numerically
\beq
{\cal A}(\bar B_s^0  \to \phi \eta )^{\rm NLO} &=& \underbrace{(0.47+9.47\; i)}_{{\cal A}( \phi\eta_q )} \cdot
\cos(39.3^\circ)
-\underbrace{(-14.76-5.76\; i)}_{{\cal A}(\phi \eta_s)} \cdot\sin(39.3^\circ)
= 9.66 +10.88\; i, \label{eq:ee21}\\
{\cal A}(\bar B_s^0  \to \phi \etar )^{\rm NLO} &=& \underbrace{(0.47+9.47\; i)}_{{\cal A}( \phi\eta_q )} \cdot \sin(39.3^\circ)
+\underbrace{(-14.76-5.76\; i)}_{{\cal A}(\phi \eta_s)} \cdot\cos(39.3^\circ)
= -11.02 +1.57\; i,\label{eq:ee22}
\eeq
and similar numerical results for ${\cal A}( B_s^0  \to \phi \eta )^{\rm NLO}$ and ${\cal A}( B_s^0  \to \phi \etar
)^{\rm NLO}$. It is then simple to define the ratio $R_5^{\rm NLO}$ in the following form
\beq
R_5^{\rm NLO}(|\cala |^2)=\frac{|{\cal A}(\bar B_s^0  \to \phi \eta )^{\rm NLO}|^2 + |{\cal A}( B_s^0  \to \phi \eta )^{\rm NLO}|^2 }{
|{\cal A}(\bar B_s^0  \to \phi \etar )^{\rm NLO}|^2 + |{\cal A}(B_s^0  \to \phi \etar )^{\rm NLO}|^2}
=2.2 \sim R_5^{\rm NLO}(\calb).
\eeq
One can see that the strength of the interference at the NLO level become a little weaker than that at the LO level,
the value of the ratio consequently  changed its value from a large $13$ to a relatively small one $ 2.2$.
For $\bar B_s^0  \to \phi \etar$ decay, based on the data collected in the period of RUN-I,
the LHCb Collaboration put an upper limit on its branching ratio very recently \cite{lhcb-d2}:
$\calb(\bar{B}_s^0 \to \phi \etar) < 0.82\times 10^{-6}$ at $90\%$ and $1.01\times 10^{-6}$ at $95\%$ confidence level (CL).
The PQCD predictions and  the QCDF prediction as given in Ref.~\cite{npb675} agree with this limit, while
the central value of the QCDF prediction as given in Ref.~\cite{chengbs09} is likely too large.
The future LHCb and/or Belle-II measurements for this kind of decays may be helpful for us to
examine the mixing pattern between $\eta-\etar$ system.

For $\bar B_s^0  \to \omega \eta$ and $ \omega \etar $  decays, the difference between the PQCD predictions for
their branching ratios can be understood by a similar mechanism: the interference effects between the decay amplitude
$\cala(\omega \eta_q)$ and $\cala(\omega \eta_s)$.

\item[(5)]
For the ``color-suppressed-tree" decay $\bar B_s^0  \to \eta K^{*0}$, the total NLO contribution is negligibly small.
For other three same kind decays $\bar B_s^0  \to K^0 (\rho^0, \omega)$
and $\bar B_s^0  \to \etar K^{*0}$,
however, the NLO contributions can provide a factor of $2-4$ enhancement to their branching ratios.
The central values of the NLO PQCD predictions for the branching ratios of above four decays agree well
with those QCDF predictions as given in Ref.~\cite{npb675} within one standard deviation,
but smaller than those QCDF predictions as given in Ref.~\cite{chengbs09} by a factor of $3-5$.
Such model differences will be examined by the LHCb (RUN-II) and/or Belle-II experiments.

\item[(6)]
For the three ``Electroweak-Penguin" $\bar B_s^0  \to \etap \rho^0$ and $\pi^0\phi $ decays,
the NLO corrections comes only from the usage of the NLO Wilson coefficients $C_i(\mu)$, the $\alpha_s(\mu)$ at
two-loop level  and  the so-called `` Vertex corrections".  The enhancement or reduction
due to the inclusion of the NLO contributions are always not large: less than $45\%$ in magnitude.
The PQCD predictions for their decay rates agree well with those in QCDF approach \cite{npb675,chengbs09}.

\item[(7)]
For the four pure ``annihilation" decays, the only NLO correction comes from the usage of the NLO Wilson
coefficients $C_i(\mu)$ and the $\alpha_s(\mu)$ at two-loop level.
For $\bar{B}_s \to \pi^\mp \rho^\pm$ and $\pi^0 \rho^0$ decays, the NLO corrections
will lead to $\sim 50\%$ reduction on their LO PQCD predictions for branching ratios, but the NLO PQCD
predictions are still at the $10^{-7}$ level, much larger than those QCDF predictions as given in
Refs.~\cite{npb675,chengbs09} by roughly one to two orders of magnitude.
The forthcoming LHCb and Belle II experimental measurements can help us to examine such large
theoretical difference.

The $\bar{B}_s \to \pi^0 \omega$ decay is also a pure  ``annihilation"  decay, but the theoretical
predictions for its branching ratios in both the PQCD and QCDF approaches are always tiny in size:
less than $10^{-8}$ and be hardly measured even in the future LHCb experiments.

\item[(8)]
By comparing the numerical results as listed in the sixth column (``+MP") and seventh column (``NLO"),
one can see easily that, the effects due to the inclusion of the NLO pieces of the $B_s \to K$ or $B_s \to \eta_s$
transition form factors are always small: $\sim 10\%$ for the first seventeen decays.
For the remaining four pure ``annihilation" decays, in fact, they do not receiver such kinds of NLO corrections.

\item[(9)]
The still missing NLO contributions in the pQCD approach are the ones to
the LO hard spectator and the non-factorizable annihilation diagrams. But from the comparative studies
for the LO and NLO contributions from different sources in Refs.~\cite{xiao2012,xiao2014,npb896-355},
we do believe that those still missing NLO contributions are most possibly the higher order corrections to the small LO
quantities, and therefore can be safely neglected.

\end{itemize}

Now we turn to the evaluations of the CP-violating asymmetries for the considered decay modes.
In the  $B_s$ system, we expect a much larger decay width difference: $\Delta\Gamma_s/(2\Gamma_s)\sim
-10\% $ \cite{pdg2016}. Besides the direct CP violation $\cala_f^{dir}$, the CP-violating asymmetry
$S_f$ and $H_f$ are defined as usual \cite{ali07}
 \beq
\cala_f^{\rm dir}=\frac{|\lambda|^2-1 }{1+|\lambda|^2},\quad
{\cal S}_f=\frac{2 {\rm Im}[\lambda]}{1+|\lambda|^2},\quad
\calh_f=\frac{2 {\rm Re}[\lambda]}{1+|\lambda|^2}.
\eeq
They satisfy the normalization relation $|\cala_f|^2+|{\cal S}_f|^2+|\calh_f|^2=1$, while
the parameter $\lambda$ is of the form
\beq
\lambda=\eta_f e^{2i\beta_s}\frac{A(\ov B^0_s \to f)}{A(B^0_s \to \bar f)},
\eeq
where $\eta_f$ is $+1(-1)$ for a CP-even(CP-odd) final state $f$ and $\beta_s
=\arg[-V_{ts}V_{tb}^*]$ is very small in size.

The PQCD predictions for the direct CP asymmetries $\cala_f^{\rm dir}$, the mixing-induced CP
asymmetries $S_f$ and $H_f$ of the considered decay modes are listed in Table
\ref{tab:acp1} and Table \ref{tab:acp2}. In these two tables, the label ``LO" means the
LO PQCD predictions, the label ``NLO" means that all currently
known NLO contributions are taken into account, the same definition as for the NLO PQCD predictions
for the branching ratios as in Table \ref{tab:br1}. The errors here are defined in the same way
as for the branching ratios.
As a comparison, the LO PQCD predictions as given in Ref.~\cite{ali07}
and the central values of the NLO QCDF predictions as given in Ref.~\cite{npb675} are also listed in
Table \ref{tab:acp1} and \ref{tab:acp2}. Since the mechanism and the sources of the CP asymmetries
for the considered decay modes are very different in the PQCD approach and the QCDF approach, we here
listed the central values of the NLO QCDF predictions only.
Unfortunately, no experimental measurements for the CP asymmetries of the $B^0_s$ decays considered here
are available at present.

\begin{table}
\caption{ The LO and NLO PQCD predictions for the direct CP asymmetries  $\cala_f^{\rm dir}$ (in units of $10^{-2}$ )
of the considered $\bar B_s^0 \to PV $ decays. As comparisons, the LO PQCD predictions as given in Ref.~\cite{ali07}
and the central values of the NLO QCDF predictions as given in Ref.~\cite{npb675} are listed in last two columns. }
 \label{tab:acp1}
\begin{tabular*}{10cm}{@{\extracolsep{\fill}}l|l|ll|ll} \hline\hline
{\rm Mode}&{\rm Class} &{\rm LO}&{\rm NLO}&{\rm PQCD}\cite{ali07}&{\rm QCDF}\cite{npb675} \\ \hline
$\bar B_s^0\to K^+\rho^-$&      {\rm T}&    $16.5 $  &$11.3^{+2.9}_{-2.8}        $&$14.2^{+3.5}_{-5.6} $&$-1.5$\\
$\bar B_s^0\to \pi^{0}K^{*0}$&  {\rm C}&    $-49.4  $&$19.7^{+3.7}_{-4.9}       $&$-47.1^{+36.4}_{-31.8}   $&$-45.7$\\
$\bar B_s^0\to K^0 {\rho}^{0}$& {\rm C}&   $72.1   $&$69.4^{+6.2}_{-5.5} $&$73.4^{+17.5}_{-49.4}   $&$24.7$\\
$\bar B_s^0\to K^0 \omega$&     {\rm C}&   $-59.3  $&$-84.7^{+1.1}_{-4.5}    $&$-52.1^{+23.1}_{-15.2}   $&$-43.9$\\
$\bar B_s^0\to \pi^0\phi$&     {\rm P$_{\rm EW}$}& $15.1   $&$49.2^{+0.4}_{-0.5}       $&$13.3^{+2.6}_{-1.8} $&$27.2$\\
$\bar B_s^0\to K^0\phi$&       {\rm P}&     $ -       $&$-2.9^{+1.2}_{-1.4}       $&$0 $&$-10.3 $\\ \hline
$\bar B_s^0\to \pi^0 \omega$&  {\rm ann}&     $4.7     $ &$3.8^{+0.5}_{-0.7}      $&$6.0^{+0.9}_{-6.2}  $&$- $\\
$\bar B_s^0\to \pi^- \rho^+$&   {\rm ann}&     $1.6   $&$-1.2^{+3.2}_{-2.3}         $&$4.6^{+2.9}_{-3.6} $&$-$\\
$\bar B_s^0\to \pi^+\rho^-$&   {\rm ann}&     $-4.3  $&$-8.5^{+5.7}_{-4.8}     $&$-1.3^{+2.9}_{-3.5}    $&$- $\\
$\bar B_s^0\to \pi^0\rho^0$&   {\rm ann}&     $1.0    $&$4.6^{+2.5}_{-3.6}        $&$1.7^{+3.9}_{-3.6}    $&$-  $\\ \hline
$\bar B_s^0\to \pi^-K^{*+}$&   {\rm T}&     $-17.2  $& $-12.1^{+1.2}_{-3.5}     $&$-19.0^{+3.7}_{-5.6} $&$0.6 $\\
$\bar B_s^0\to K^+K^{*-}  $&   {\rm P}&     $-34.1    $&$-21.6^{+4.9}_{-4.3}     $&$-36.6^{+3.8}_{-4.3}   $&$2.2 $\\
$\bar B_s^0\to K^- K^{*+} $&       {\rm P}&    $50.1    $& $46.6^{+7.4}_{-6.7}       $&$55.3^{+10.8}_{-11.2}   $&$-3.1 $\\
$\bar B_s^0\to K^{0}\bar K^{*0}$& {\rm P}&    $-  $& $0.8^{+0.1}_{-0.1}       $&$0   $&$1.7 $\\
$\bar B_s^0\to \bar{K}^0K^{*0}$&   {\rm P}&    $-  $ & $0.1^{+0.05}_{-0.05}       $&$0   $&$0.2$\\ \hline
$\bar B_s^0\to \eta K^{*0}$&       {\rm C}&    $38.5  $& $30.6^{+10.9}_{-8.5}       $&$51.2^{+15.6}_{-14.4}   $&$40.2$\\
$\bar B_s^0\to \eta^\prime K^{*0}$&     {\rm C}&      $-37.2   $&$-63.4^{+3.1}_{-2.6}   $&$-51.1^{+16.1}_{-19.8}   $&$-58.6$\\
$\bar B_s^0\to \eta {\rho}^{0}$&          {\rm P$_{\rm EW}$}&  $-14.3  $&$37.7^{+1.2}_{-2.3}       $&$-9.2^{+3.1}_{-2.8}  $&$27.8$\\
$\bar B_s^0\to \eta^\prime {\rho}^{0}$& {\rm P$_{\rm EW}$}&      $23.9 $& $54.1^{+1.2}_{-1.3}       $&$25.8^{+4.6}_{-4.4}  $&$28.9$\\
$\bar B_s^0\to  \eta \omega $&         {\rm P, C}&         $-9.4   $&$-35.8^{+2.3}_{-3.5}        $&$-16.7^{+16.5}_{-19.4} $&$-$\\
$\bar B_s^0\to  \eta^\prime \omega  $ &{\rm P, C}&         $11.3   $&$-23.5^{+5.4}_{-4.6}        $&$7.7^{+10.4}_{-4.2}    $&$- $\\
$\bar B_s^0\to \eta \phi $&           {\rm P}&         $-1.2  $&$-4.1^{+0.2}_{-0.3}   $&$-1.8^{+0.6}_{-0.6}    $&$-8.4$\\
$\bar B_s^0\to  \eta^\prime \phi$&    {\rm P}&         $5.1  $& $14.2^{+1.3}_{-2.5}          $&$7.8^{+1.9}_{-8.6} $&$-62.2$\\ \hline\hline
\end{tabular*} \end{table}

\begin{table}
\caption{The LO and NLO PQCD predictions for the mixing-induced CP asymmetries (in units of $10^{-2}$)
$S_f$ ( the first row) and $H_f$ (the second row). The meaning of the labels are the same as those
in Table~\ref{tab:acp1}. }
\label{tab:acp2}
\begin{tabular*}{8cm}{@{\extracolsep{\fill}}l|l|ll|l} \hline\hline
 {\rm Mode}  & {\rm Class } & {\rm LO} &  {\rm NLO}&{\rm PQCD} \cite{ali07} \\ \hline
$\bar B_s^0\to K_S {\rho}^0 $   &{\rm C}&$-54.1 $  &$-7.9^{+15.2}_{-16.2}$&$-57^{+56}_{-43}$\\
                                      &  &$-41.2 $  &$-71.5^{+8.9}_{-5.4} $&$-36^{+47}_{-20}$\\
$\bar B_s^0\to K_{S}\; \omega$  &{\rm C}&$-55.5 $  &$30.1^{+5.2}_{-8.4} $&$-63^{+29}_{-14}$\\
                                       & &$-58.3 $  &$-43.8^{+5.6}_{-3.2} $&$-57^{+33}_{-40}$\\
$\bar B_s^0\to K_{S}\; \phi$    & {\rm P}          &$-72.1 $  &$-95.6^{+0.2}_{-0.1} $&$-72$\\
                                        & &$-69.3$  &$-33.4^{+2.2}_{-1.0}$&$-69$\\ \hline
$\bar B_s^0\to \pi^0 \phi $     &{\rm P$_{\rm EW}$} &$-16.1 $  &$-8.7^{+4.5}_{-4.3} $&$-7^{+8}_{-10}$\\
                                        &&$97.2 $   &$86.6^{+0.6}_{-0.8}$&$98^{+1}_{-3}$\\
$\bar B_s^0\to \pi^0 \rho^0 $   & {\rm Anni}        &$-20.1  $ &$-24.5^{+1.4}_{-2.9}  $&$-19^{+2}_{-3}$\\
                                       & &$97.3$    &$96.8^{+0.3}_{-0.7}$&$99$\\
$\bar B_s^0\to \pi^0 \omega $  & {\rm Anni}      &$-97.2  $ &$-97.6^{+0.1}_{-0.2}  $&$-97^{+11}_{-2}$\\
                                       & &$-24.6$   &$-21.2^{+1.1}_{-0.4}$&$-22^{+13}_{-29}$\\ \hline
$\bar B_s^0\to \eta \omega $     &{\rm P, C}        &$-4.5 $  &$11.1^{+2.3}_{-2.1}  $&$-2^{+2}_{-9}$\\
                                       & &$97.5 $  &$93.3^{+0.4}_{-0.3} $&$99^{+1}_{-6}$\\
$\bar B_s^0\to \eta^{\prime} \omega$ &{\rm P, C}    &$-19.3$    &$-35.4^{+5.2}_{-7.9}$&$-11^{+5}_{-5}$\\
                                       & &$96.6 $   &$91.8^{+3.3}_{-4.1} $&$99 $\\
$\bar B_s^0\to \eta {\rho}^{0} $ &{\rm P$_{\rm EW}$}        &$20.8 $  &$10.3^{+2.3}_{-2.6} $&$15^{+15}_{-17} $\\
                                       &  &$97.1 $  &$92.6^{+0.4}_{-0.1}$&$98^{+1}_{-3}$\\
$\bar B_s^0\to \eta^{\prime} \rho^0$ &{\rm P$_{\rm EW}$} &$-29.4 $ &$-11.1^{+1.3}_{-1.6}  $&$-16^{+11}_{13} $\\
                                       &  &$92.2$   &$82.3^{+3.6}_{-3.7}$&$95^{+1}_{-3}$\\
$\bar B_s^0\to \eta {\phi}          $   &{\rm P}    &$-3.2  $ &$-4.2^{+0.5}_{-0.5}  $&$-3^{+7}_{-21}$\\
                                       &  &$99.9$   &$99.4^{+0.1}_{-0.1}$&$100^{+0}_{-1}$\\
$\bar B_s^0\to \eta^{\prime} \phi  $  &{\rm P}     &$-8.6  $ &$-6.1^{+0.6}_{-0.4}   $&$0^{+2}_{-2}$\\
                                       &  &$99.9$   &$99.1^{+0.1}_{-0.1}$&$100^{+0}_{-2}$\\
 \hline\hline
\end{tabular*} \end{table}

From the PQCD predictions for the CP violating asymmetries of the
considered $\bar{B}^0_s$ decays as listed  in the Table~\ref{tab:acp1} and \ref{tab:acp2}, one can see the
following points:
\begin{itemize}
\item[(1)]
For all $\bar B_s^0\to PV$ decays, the LO PQCD predictions for their CP asymmetries
obtained in this paper do agree well with those as given in Ref.~\cite{ali07}.

\item[(2)]
For most $\bar B_s^0\to PV$ decays, the changes of the PQCD predictions for the CP asymmetries
induced by the inclusion of the NLO corrections are basically not large in size.
For $\bar B_s^0\to \pi^0 K^{*0}$, $\eta \rho^0$ and $\etar \omega$ decays, however, the PQCD predictions
for their $\cala_f^{dir}$ can change sign after the inclusion of the NLO corrections.
For $\bar B_s^0\to \pi^0 \phi$, $\etar \rho^0 $, $\eta\omega$ and $\etar \phi$ decays, on the other hand,
the NLO enhancements on their $\cala_f^{dir}$ can be larger than a factor of two.

\item[(3)]
By comparing the numerical results as listed in Table \ref{tab:acp1},
one can see that the PQCD and QCDF predictions for the CP-asymmetries of the considered decays are indeed quite
different, due to the very large difference in the mechanism to induce the CP asymmetries in the pQCD approach and the
QCDF approach.
In the PQCD approach, fortunately, one can calculate the CP asymmetries for the pure annihilation decays.
From Table \ref{tab:acp1} one can see that the PQCD predictions for the  $\cala_f^{dir}$ of the four
pure annihilation decays $\bar B_s^0\to \pi (\omega, \rho) $ are small: less than $10\% $ in magnitude.

\item[(4)]
Since the currently measured $\bar{B}_s \to \pi^- K^{*+} $ and $\bar{B}_s \to K^+ K^{*-} + K^- K^{*+} $ decays
have a large decay rates at the level of $10^{-5} - 10^{-6}$,
their relatively large direct CP asymmetries from $-30\%$ to around $50\%$ could be measured in the near future
LHCb or Belle-II experiments.
For  $\bar{B}_s \to K^0(\rho^0,\omega)$ and $\etar (K^{*0},\rho^0)$, however, it might be very
difficult to measure their large direct CP asymmetries ( around $50\%$ in magnitude ), due to their very
small branching ratios at the level of $10^{-7} - 10^{-8}$.

\item[(5)]
The mixing-induced $CP$ asymmetries $S_f$ and $H_f$ for the considered twelve decay modes are shown in
Table \ref{tab:acp2}. For $\bar{B}_s \to K_s (\omega,\phi)$ and $ \omega(\pi^0, \etar)$ decays, although
their $S_f$ are large in size, but it is still very difficult to measure them due to their very small
decay rates.

\end{itemize}

\section{SUMMARY}\label{sec:4}

In summary, we calculated the CP-averaged branching ratios and CP-violating asymmetries for all twenty one
$\bar B_s^0\to P V$ decays with $P=(\pi, K, \eta, \eta^\prime)$ and $V=(\rho, K^*, \phi,\omega)$
by employing the PQCD factorization approach.
All currently known NLO contributions, specifically those newly known NLO twist-2 and twist-3
contributions to the relevant form factor $F_0^{B^0_s\to K}(0)$ and $F_0^{B^0_s\to \eta_s}(0)$,
are taken into account.

From our analytical evaluations and numerical calculations, we found the following points:
\begin{itemize}
\item[(1)]
The LO PQCD predictions for the branching ratios and CP-violating asymmetries of
$B_s \to PV$ decays as presented in Ref.\cite{ali07} are confirmed by our independent calculations.
The effects of the NLO contributions on the PQCD predictions for the branching ratios and CP asymmetries
of the considered decay modes are channel dependent and will be tested by future experiments.

\item[(2)]
For the three measured decays $\bar{B}_s^0  \to K^0 \bar{K}^{*0}$, $K^{\pm}K^{*\mp}$
and $\pi^- K^{*+}$, the NLO contributions can provide a large enhancement ( about $30-45\%$) or
a reduction ($\sim 37\%$) to the LO PQCD predictions for their branching ratios, respectively.
From the variations of the ratios $R_{1,2,3}$, one can see that the agreements between the PQCD predictions
and the measured values are improved significantly due to the inclusion of the NLO contributions.
This is the major reason why we have made great efforts to calculate the NLO contributions in the PQCD
factorization approach.

\item[(3)]
For the considered $B_s \to PV$ decays, the effects from the inclusion of the
NLO twist-2 and twist-3 contributions to the form factor  $F_0^{B^0_s\to K}$ and $F_0^{B^0_s\to \eta_s}$
are always small:  less than $ 10\%$ in magnitude.

\item[(4)]
For the ``tree" dominated decay $\bar B_s^0\to K^+ \rho^-$ and the ``color-suppressed-tree" decay
$\bar B_s^0\to \pi^0 K^{*0}$ decay, the different topological structure and the strong
interference effects (constructive or destructive) between decay amplitude
$\cala_{T,C}$ and $\cala_{P}$  together leads to the very large difference in their decay rates.

\item[(5)]
For $\bar{B}_s^0 \to V(\eta, \etar)$ decays, the complex pattern of the PQCD predictions for their
branching ratios can be understood by the difference of the major contributing Feynman diagrams, and the
interference effects (constructive or destructive) between the decay amplitude $\cala(V\eta_q)$
and $\cala(V\eta_s)$ due to the $\eta-\etar$ mixing.

\end{itemize}

\begin{acknowledgments}

This work is supported by the National Natural Science
Foundation of China under Grants  No.~11775117, 11235005  and 11205072,
and by the Research Fund of Jiangsu Normal University under Grant No.~HB2016004.

\end{acknowledgments}



\end{document}